\documentclass[journal,twoside]{IEEEtran}

\usepackage{generic}
\usepackage{cite}
\usepackage{amsmath,amssymb,amsfonts}
\usepackage{algorithmic}
\usepackage{graphicx}
\usepackage{algorithm,algorithmic}
\usepackage{hyperref}
\hypersetup{hidelinks=true}
\usepackage{textcomp}
\usepackage{subfigure}
\usepackage{adjustbox}
\usepackage{graphicx}
\usepackage{array}
\usepackage{multirow}
\usepackage{tabularx}

\begin{document}
\title{Unsupervised Hybrid framework for ANomaly Detection (HAND) - applied to Screening Mammogram}
\author{Zhemin Zhang, Bhavika Patel, Bhavik Patel, Imon Banerjee %
\thanks{Zhemin Zhang is with Department of Computer Science, Arizoan State University}%
\thanks{Bhavika Patel, Bhavik Patel, Imon Banerjee are with the Department
of Radiology, Mayo Clinic, Arizona}
}

\markboth{IEEE JOURNALS OF BIOMEDICAL AND HEALTH INFORMATICS}%
{Zhang \MakeLowercase{\textit{et al.}}: Unsupervised Hybrid framework for ANomaly Detection (HAND) - applied to Screening Mammogram}

% \begin{document}
% <-this % stops a space
% <-this % stops a space
% \IEEEmembership{Member, IEEE}
% \thanks{This paragraph of the first footnote will contain the date on 
% which you submitted your paper for review. It will also contain support 
% information, including sponsor and financial support acknowledgment. For 
% example, ``This work was supported in part by the U.S. Department of 
% Commerce under Grant 123456.'' }
% \thanks{The next few paragraphs should contain 
% the authors' current affiliations, including current address and e-mail. For 
% example, First A. Author is with the National Institute of Standards and 
% Technology, Boulder, CO 80305 USA (e-mail: author@boulder.nist.gov). }
% \thanks{Second B. Author Jr. was with Rice University, Houston, TX 77005 USA. He is 
% now with the Department of Physics, Colorado State University, Fort Collins, 
% CO 80523 USA (e-mail: author@lamar.colostate.edu).}
% \thanks{Third C. Author is with 
% the Electrical Engineering Department, University of Colorado, Boulder, CO 
% 80309 USA, on leave from the National Research Institute for Metals, 
% Tsukuba, Japan (e-mail: author@nrim.go.jp).}}

\maketitle

\begin{abstract}
Out-of-distribution (OOD) detection is crucial for enhancing the generalization of AI models used in mammogram screening. Given the challenge of limited prior knowledge about OOD samples in external datasets, unsupervised generative learning is a preferable solution which trains the model to discern the normal characteristics of in-distribution (ID) data. The hypothesis is that during inference, the model aims to reconstruct ID samples accurately, while OOD samples exhibit poorer reconstruction due to their divergence from normality. Inspired by state-of-the-art (SOTA) hybrid architectures combining CNNs and transformers, we developed a novel backbone - HAND, for detecting OOD from large-scale digital screening mammogram studies. To boost the learning efficiency, we incorporated synthetic OOD samples and a parallel discriminator in the latent space to distinguish between ID and OOD samples. Gradient reversal to the OOD reconstruction loss penalizes the model for learning OOD reconstructions. An anomaly score is computed by weighting the reconstruction and discriminator loss. On internal RSNA mammogram held-out test and external Mayo clinic hand-curated dataset, the proposed HAND model outperformed encoder-based and GAN-based baselines, and interestingly, it also outperformed the hybrid CNN+transformer baselines. Therefore, the proposed HAND pipeline offers an automated efficient computational solution for domain-specific quality checks in external screening mammograms, yielding actionable insights without direct exposure to the private medical imaging data.
\end{abstract}

\begin{IEEEkeywords}
OOD detection, Unsupervised learning, CNN+Transformer, Screening Mammograms
\end{IEEEkeywords}

\section{Introduction}\label{sec1}
\label{sec:introduction}
\IEEEPARstart{I}{n} medical imaging domain, supervised deep learning has been promising in solving various tasks, but it requires well-annotated datasets for training and validation which must be extracted and curated to a expert level quality to ensure superior performance~\cite{van2019quality}. However, the primary challenge is that the medical datasets from different institutions can be of heterogeneous in nature due to difference in equipment, acquisition protocol, and population drift, resulting in data distribution shifts between sets. Distribution shift is a pervasive issue in predictive modeling across practical applications, often arising from biases introduced by experimental design and discrepancies in testing conditions compared to training \cite{quinonero2009dataset}. Imbalanced data, domain shift, and source component shift are among the most prevalent forms of this challenge \cite{storkey2009training}. The shift data is considered out-of-distribution (OOD) in the dataset, and should account for the performance dropping of well-trained models. Thus, identifying the shift data (OODs) is crucial for cleaning the datasets and helpful in enhancing the model's generalization with future training. 

According to USPSTF new guideline recommendations for breast cancer screening, women should start screening mammograms at 40 years old which resulted a huge number of exams for interpretation by radiologists.~\cite{berg2024uspstf}. Mammography has a number of known limitations, including imperfect sensitivity and specificity and inter-reader variability which increase radiology's workload and reading time~\cite{abdelhafiz2019deep}. To increase the reading efficiency, currently there are 882 FDA approved AI/ML-Enabled Medical Devices and among them 18 are specifically for breast cancer screening\footnote{https://www.fda.gov/medical-devices/software-medical-device-samd/artificial-intelligence-and-machine-learning-aiml-enabled-medical-devices}. Despite the available software solution, generalization to an unseen population is a crucial for such AI deployment in clinical workflow; however due to the artifacts and quality variations in mammogram, AI generalization can be a extremely challenging problem. ODD detection can be a potential solution which can identify the anomaly samples from the unseen distribution of the screening images for expert review and able to significantly boost the model generalization performance on the in-distribution cases (ID). 

\begin{figure}[htb!]
\centering
\begin{subfigure}[]{}
\includegraphics[width=0.3\linewidth]{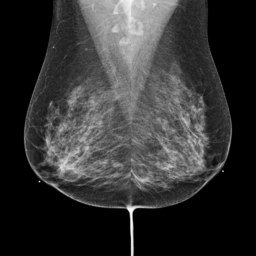}
    % \caption{ID example}
    \label{fig:id_example}
\end{subfigure}
\begin{subfigure}[]{} \includegraphics[width=0.3\linewidth]{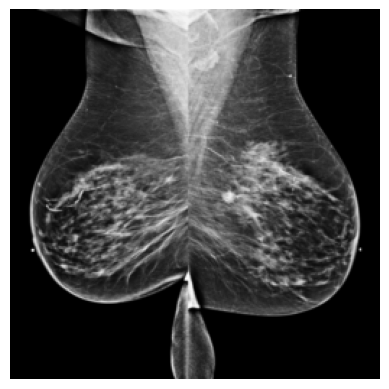}
    \label{fig:bc_example}
\end{subfigure}
\begin{subfigure}[]{}\includegraphics[width=0.3\linewidth]{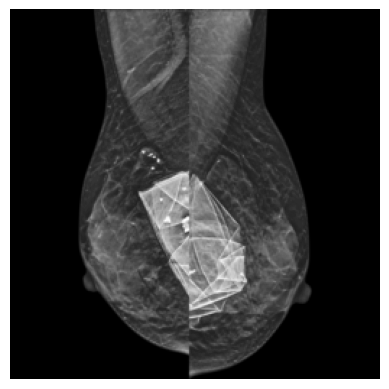}
    \label{fig:implant_example}
\end{subfigure}
\begin{subfigure}[]{} \includegraphics[width=0.3\linewidth]{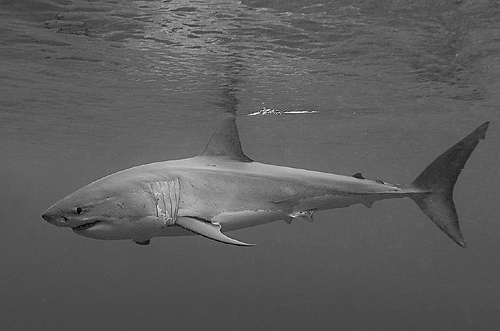}
    \label{fig:implant_example}
\end{subfigure}
\caption{Comparative examples between ID, intra-class and inter-class - (a) ID mammogram. (b) Intra-class OOD example is biopsy clip and (c) is implant rapture, which share significant similarity with ID mammogram. (d) Inter-class OOD example from ImageNet, which is significantly different from mammogram images.}
\label{fig:id_ood_example}
\end{figure}

\textit{OOD} data, also called \textit{anomaly}, \textit{outlier}, usually refers to data that shows dissimilarity from the training or in distribution (ID). Given an image $x$, the goal of \textit{OOD detection} is to identify whether $x$ is from ID dataset $D_{in}$ or OOD dataset $D_{out}$. 
OOD data can be categorized as two broad classes (Fig.~\ref{fig:id_ood_example}) - \textit{(i) intra-class:} OOD data belonging this type, which is also called \textit{novelty data}, often shares severe similarity with the ID classes and is extremely challenging to distinguish, \textit{e.g.}, the mammogram with biopsy clip presents close appearance with the normal images; \textit{(ii) inter-class:} this data is significantly different from ID samples, \textit{e.g.,} a natural image from ImageNet is much different in shape and color from the mammogram images. In theory, there exists an infinite number of potential categories for intra- and inter-class OOD scenarios within each domain. This complexity makes training a supervised model for OOD identification impractical, especially when considering unseen external data. 

Within the scope of this article, we developed an unsupervised \textbf{H}ybrid \textbf{AN}omaly \textbf{D}etection framework (HAND) by combining CNN + VisionTransformer (ViT). While CNN architecture is known for being capable to capture local spatial features,  the transformer in ViT has strength in capturing the long-range dependencies by using self-attention mechanism. Thus ViT and CNN have complimentary strengths to process medical imaging for various applications, and many works have been attempting to combine these two architectures to have strength to capture both global and local contents of the images as hybrid concept architectures~\cite{kim2024systematic}.

Within the scope of this study, our primary contributions for devising \emph{HAND} are:

(i) We extended the TransBTS hybrid architecture~\cite{li2022transbtsv2} with a discriminator branch and gradient reversal to support efficient detection of intra- and inter-class of anomaly samples. 

(ii) Devise an synthetic ODD generation technique leveraging geometric augmentation to support unsupervised training of the model with penalization. 

(iii) Train the model using only ID data from RSNA Screening Mammography Breast Cancer Detection AI Challenge (2023) dataset\footnote{https://www.rsna.org/rsnai/ai-image-challenge/screening-mammography-breast-cancer-detection-ai-challenge} and validate the model on OOD data from RSNA internal dataset and unseen Mayo clinic dataset.

\section{Related Work}\label{sec2}
Though there is extensive research on OOD detection, there is a need for more targeted unsupervised development needed in medical domain to boost the AI model generalization~\cite{Tharindu2022survey}. We have divided unsupervised OOD detection methods into two main categories - classifier-base and generative method. \\

\noindent \emph{Classifier-base anomaly detection: } Anomaly/ODD detection problem can be solved by one-class-classifier (OCC)~\cite{Khan2013OCC} when the OOD samples is absent or not well defined, such as DeepSVDD~\cite{ruff2018deepOCC}, FCDD~\cite{Liznerski2020FCDD}, DOC~\cite{Perera2018DOC}, and OC-SVM~\cite{Bernhard2001ocsvm}. OCC approaches generally has access to only ID-data for training, and optimize a kernel-base objective function to learn a hyper-sphere separating ID and OOD samples. This approach does not perform well in intra-class OOD which share more similarity with ID data. ODIN~\cite{Liang2017ODIN} use temperature scaling to score function and adding perturbations to input to separate ID and OOD data. However, this approach has limited applicability in healthcare since OOD classes can be multiple, beyond what can be represented by simple perturbations to input ID data. \\

\noindent \emph{Generative anomaly detection: } Generative models are the most promising method for unsupervised OOD detection. This is accomplished by to discern the normal characteristics of in-distribution (ID) training data. Within the generative model family, variational autoencoder (VAE) is an autoencoder architecture which regularize the latent space distribution and thus enable generative ability. However, the quality of generative images are usually blurry and missing detailed information. Based on the VAE architecture, multiple studies,~\cite{zimmerer2019vae,Sakurada2014AnomalyDU,Zhou2017AnomalyDW,Beggel2020AE,An2015VariationalAB,bao2017vae} used only reconstruction error as basis of anomaly score. However, often only reconstruction error from VAE may suffer to identify complex intra-class OOD data in the mammogram domain due to the poor to moderate understanding of 'normality'. Bayesian VAE~\cite{Daxberger2019BayesianVA}~\cite{xiao2020likelihood}~\cite{RAN2022199}~\cite{Pol2019AnomalyDW}~\cite{wu2020VAE} detects OOD by estimating a full posterior distribution over decoder parameters using stochastic gradient Markov chain Monte Carlo. However, most of VAE-based OOD detection are only evaluated on the natural images, which bear a clear separation for intra and inter class samples. 
While CVAD~\cite{guo2022cvad} extend the VAE architecture for medical data (chest X-ray) by combining low and high-level features to have comparable better image quality and a discriminator loss as part of anomaly score. However, like other VAE models, posterior collapse problem exists in this model and the discriminator is hard to converge to identify all class of OOD. 

Similar to VAE, GANs can learn latent feature representations by training generator to produce fake images and discriminator to distinguish real vs fake images (output by generator). Thus, discriminator latent features can be used to separate ID and OOD. But usually it is difficult to stabilize the adversarial GAN training. GANomaly~\cite{akcay2019ganomaly}, ADGAN~\cite{Lucas2019adgan}, AnoGAN~\cite{Schlegl2017anogan}, f-AnoGAN~\cite{schlegl2019f}, and OCGAN~\cite{Perera2019ocgan} are popular models for OOD detection task. However, the experiments are done using natural images or small patches of medical data, may fail to generalize with large dimensional medical images. 

Transformer excels VAE and GAN in capturing long-range dependencies which makes it suitable for OOD detection, since model can gather information of entire image to decide if it's OOD or not. For OOD detection, transformer is usually combined with CNN architecture - either transformer used as encoder and CNN as decoder, or vise-versa. Ano-ViT~\cite{leeanovit2022} and VT-ADL~\cite{pankajvtadl2021} use transformer as encoder and CNN decoder for anomaly detection. While ADTR~\cite{you2023adtr} and UTRAD~\cite{CHEN202253} apply pre-trained CNN to obtain latent representation then feed into transformer which serve as decoder. Interestingly, studies in~\cite{pateltransformerpet2022, pinayabraintrans2022}  used a hybrid architecture where Vector Quantized Variational Autoencoder (VQ-VAE) is applied to get a discrete compact latent representation of images, then applying transformer from the latent representation to identify patches and images under threshold as anomaly. But the implementation details of transformer is clear in these two studies, also an arbitrary threshold is not preferable for different datasets. For OOD detection, self-supervised learning with transformer~\cite{tianhybridmed2024} \cite{bozorgtabaramae2023}~\cite{parkssl2021}~\cite{patelssl2023} is performed via masking or adding synthetic anomaly during training. They primarily identified OOD by the difference between input and output reconstruction without discriminator prediction. 

\section{Method}\label{sec3}
\begin{figure*}[htb!]
\centering
\includegraphics[width=0.8\textwidth]{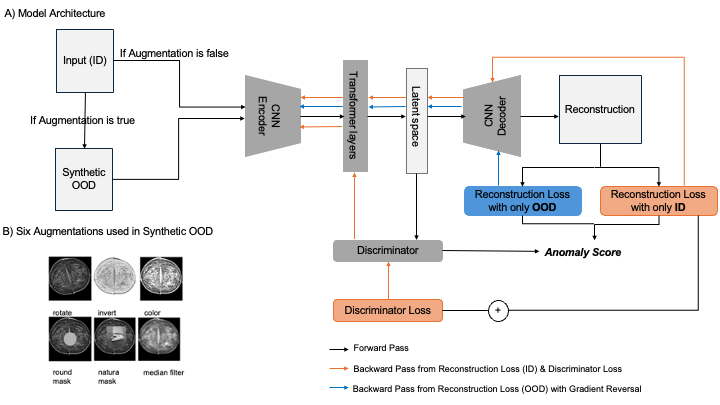} % Use \textwidth for full page width
\caption{\emph{HAND}: Proposed hybrid model framework with synthetic ODD generation: A) model architecture by combining CNN and transformer backbone; B) six different transformations for synthetic OOD generation.}
\label{fig:ModelArchitecture1}
\end{figure*}

Figure \ref{fig:ModelArchitecture1} shows design of proposed hybrid \emph{HAND}\footnote{GitHub implantation can be found: \url{https://github.com/zheminzhang96/HAND_mammo.git} } architecture for unsupervised anomaly detection which consists of a CNN-transformer backbone to learn reconstructions of input, and a parallel discriminator branch out from latent space to learn differentiation between the ID and OOD samples. HAND uses self-supervised learning (SSL) to train the hybrid architecture and the discriminator branch. Gradient reversal is applied to synthetic OOD reconstruction for model unlearning. During inference, a weighted combination of reconstruction loss and probability from discriminator branch yields anomaly score for the input. 

%\begin{verbatim}

%\end{verbatim}

\subsection{Synthetic OOD Generation for SSL training}\label{subsec2}
Given the unsupervised nature of the framework, synthetic OOD samples are generated as a part of SSL to enhance the model's ability to learn ID data. Figure \ref{fig:ModelArchitecture1} illustrates six augmentations from an input that we adopted to replicate unseen OOD appearances for the mammogram images based on limited prior knowledge. We selected color jittering, invert function and median filter to simulate the artifacts, motion and image normalization errors. Random masked images using round and natural image masks are used to replicate the possible appearances of foreign body, e.g. implant, biopsy clip. Our hypothesis is that although synthetic out-of-distribution (OOD) samples do not provide supervised training data, their close resemblance to input images (ID) will enable the model to learn subtle differences that indicate potential OODs from IDs. During model training, around half of the ID input images are selected for synthetic ODD generation by a random binary seed value. The type of augmentation applied to original input is also decided randomly from six different transformation functions. 

 %\begin{figure}[]
 %\centering
 %\includegraphics[width=0.6\textwidth]{images/augmentation.png}
 %\caption{Samples of six synthetic OOD augmentations for SSL}
 %\label{fig:OODaugmentation}
 %\end{figure}

\subsection{Hybrid Model Architecture}\label{subsec2}

We extended the TransBTS architecture~\cite{li2022transbtsv2} which is originally developed for brain MR segmentation and  composed of CNN encoder with inductive bias to grasp local spatial information of input image, followed by the transformer to learn global context. We updated the original CNN decoder to reconstruct the input image rather than mask. The selection of the hybrid architecture is influenced by the fact that the latent representation has better ability compared with CNN encoder alone because the transformer's strength in capturing global information. Therefore, a parallel discriminator branch with stacked Multilayer Perceptron (MLP) layers is added from the transformer latent space to get predicted probability of anomaly and help the model to understand the ID images. During SSL training, the image with augmentation (Synthetic OOD) is labeled as 1, otherwise (ID) is labeled 0. During training, reconstruction loss calculated as MSE loss between input image and reconstructed image is used to update the CNN encoder, transformer, and CNN decoder. \\

\noindent \emph{Discriminator} plays a crucial role in anomaly detection, since the reconstruction of input image alone is not sufficient for anomaly score~\cite{guo2023medshift, banerjee2022cvad}. This is due to the fact that the reconstruction loss for low resolution, area masked, pure black, or blurred images  are intrinsically low since the model is trained to learn to reconstruct ID images, which has more complex tissue structures. The OOD data mentioned before are easier for the model to reconstruct, resulting lower reconstruction errors, which indicates lower anomaly score for these anomalous samples. 
The discriminator shown in Figure \ref{fig:ModelArchitecture1} uses MLP framework which branches out from the latent space after transformer layers. One output from MLP with sigmoid activation function to get OOD probability. Higher probability score indicates higher chance the input image is anomaly. Discriminator loss calculated by BCE loss between actual label and predicted probability is used to update CNN encoder and transformer.

\subsection{Anomaly detection} \label{subsec2}

\noindent \emph{Gradient Reversal}
The purpose of gradient reversal in our anomaly detection framework is to increase the reconstruction error for OOD data, issuing higher anomaly score. Even though adding discriminator helps the model performance significantly, we have realized that the reconstruction quality of the OOD samples didn't degrade compared to ID data and the SSIM for implant, bad quality, and natural category shown in table \ref{tab:recon_in_ablation} and \ref{tab:recon_ex_ablation}. To help the model unlearn the OOD reconstruction and force the model to produce poor reconstruction, we designed a two step training process where gradient reversal is applied only to OOD reconstruction loss. Gradient reversal process reverses the ODD loss gradient by multiplying it with a negative scalar ($-\lambda$) during the backpropagation. We implementing such layer by
defining procedures for forwardprop (identity transform), backprop (multiplying by a constant), and parameter update (nothing). \\

\noindent \emph{Network training: }Given the open ended anomaly detection problem, the hybrid anomaly detection model is trained in an self-supervised learning (SSL) manner since no real ODD data is being used in the training. All the input images are ID data, and around half of input have random transformation to get synthetic OOD data. For each batch, we generated an even mixed of ID and OOD data during the network training. Two backpropagation steps are used in each batch to update the network. The first backpropagation uses loss formulated in Eqn. \ref{eq:backprop1}. Eqn. \ref{eq:backprop1} is composed ID reconstruction loss shown in Eqn \ref{eq:mse_id} to learn reconstruction of ID images, and BCE loss from Eqn. \ref{eq:bce} to train discriminator to distinguish ID or OOD images. The two parts of losses are been weighted by \texttt{$\alpha_1$} and \texttt{$\alpha_2$} to balanced the the 1st backpropagation loss \texttt{L$_{1}$}. The second backpropagation uses gradient reversal with only OOD reconstruction loss shown in Eqn. \ref{eqn:backprop2}. The gradient reversal is done by multiplying gradient of \texttt{L$_{2}$} with ($-\lambda$), issuing \(-\lambda \frac{\delta L_2}{\delta \theta x}\) where $x$ is the input image. 

\begin{equation}
    L_1 = \alpha_1 L(x_\text{id}, x_{\text{recon}}) + \alpha_2 D(y, y_{\text{pred}})
    \label{eq:backprop1}
\end{equation}

\begin{equation}
    L(x_\text{id}, x_{\text{recon}}) = \frac{1}{n} \sum \left(x_\text{id} - x_{\text{recon}}\right)^{2} 
    \label{eq:mse_id}
\end{equation}

\begin{equation}
    D(y, y_{\text{pred}}) = y_{\text{pred}} \log y + (1-y_{\text{pred}}) \log(1-y)
    \label{eq:bce}
\end{equation}

\begin{equation}
    L_2 = L(x_\text{ood}, x_{\text{recon}}) = \frac{1}{n} \sum \left(x_\text{ood} - x_{\text{recon}}\right)^{2}
    \label{eqn:backprop2}
\end{equation}

% \begin{equation}
%     -\lambda \frac{\delta L_2}{\delta \theta x}
% \end{equation}
\noindent \emph{Anomaly score: } Anomaly score $S$ for an image is computed as \adjustbox{max width=\columnwidth}{
% \begin{equation}
    $S = 0.5* \frac{L(x, x_{\text{recon}}) - L(x, x_{\text{recon}})_\text{min}}{L(x, x_{\text{recon}})_\text{max} - L(x, x_{\text{recon}})_\text{min}} + 0.5* \frac{y_{\text{pred}} - y_{\text{pred}_{\text{min}}}} {y_{\text{pred}_\text{max}} - y_{\text{pred}_{\text{min}}}}$
    % S = 0.5 \cdot \frac{L(x, x_{\text{recon}}) - L(x, x_{\text{recon}})_{\text{min}}}{L(x, x_{\text{recon}})_{\text{max}} - L(x, x_{\text{recon}})_{\text{min}}} 
    % + 0.5 \cdot \frac{y_{\text{pred}} - y_{\text{pred}_{\text{min}}}}{y_{\text{pred}_{\text{max}}} - y_{\text{pred}_{\text{min}}}}
    % \label{eq:ano_score}
% \end{equation}
} and derived individually during inference. The score is equally weighted between reconstruction error $L(x, x_{\text{recon}})$ and discriminator score $y_\text{pred}$. Before summation of two parts together, we first normalize within the range of $[0, 1]$ to ensure balanced weight between reconstruction error and predicted probability of anomaly from discriminator. 

% \adjustbox{max width=\columnwidth}{
% % \begin{equation}
%     $S = 0.5* \frac{L(x, x_{\text{recon}}) - L(x, x_{\text{recon}})_\text{min}}{L(x, x_{\text{recon}})_\text{max} - L(x, x_{\text{recon}})_\text{min}} + 0.5* \frac{y_{\text{pred}} - y_{\text{pred}_{\text{min}}}} {y_{\text{pred}_\text{max}} - y_{\text{pred}_{\text{min}}}}$
%     % S = 0.5 \cdot \frac{L(x, x_{\text{recon}}) - L(x, x_{\text{recon}})_{\text{min}}}{L(x, x_{\text{recon}})_{\text{max}} - L(x, x_{\text{recon}})_{\text{min}}} 
%     % + 0.5 \cdot \frac{y_{\text{pred}} - y_{\text{pred}_{\text{min}}}}{y_{\text{pred}_{\text{max}}} - y_{\text{pred}_{\text{min}}}}
%     % \label{eq:ano_score}
% % \end{equation}
% }

\section{Results}\label{sec4}
\subsection{Datasets}
For the model training and internal validation, we leveraged the publicly available RSNA Screening Mammography Breast Cancer Detection AI Challenge (2023) dataset\footnote{https://www.rsna.org/rsnai/ai-image-challenge/screening-mammography-breast-cancer-detection-ai-challenge}. The dataset is consist of standard digital mammography images which was contributed by mammography screening programs in Australia and the U.S. It includes detailed diagnosis and artifact labels, with radiologists’ evaluations and follow-up pathology results for suspected malignancies. To select the training ID cases, we first select the cases without implant, biopsy clip and artifacts based on metadata description. Afterwards, we manually filter that ID cases and drop all the potential ODD samples to create a clean dataset of 7,623 studies with 11,722 images (left and right views are stitched together). A testset of 300 studies was created by manually reviewing the images and labeling implant and poor diagnostic quality. Detailed overall characteristics of the training and testing data is reported in Table~\ref{tab:dataset_stat}.

\noindent \emph{External testing:} With the approval of Mayo Clinic Institutional Review Board (IRB), we collected a limited sample of 220 de-identified OOD studies for external validation of the model to test the generalization on an unseen population since no mammogram study from Mayo Clinic was included in RSNA dataset. OOD studies were collected based on the mention of biopsy clip, motion artifacts and poor quality, and implant in the textual radiology reports, and manually reviewed the images afterwards.

\begin{table}[]
\caption{Dataset description - RSNA Screening Mammography Breast Cancer Detection AI Challenge dataset for training and internal testing; Mayo Clinic Private dataset for External testing\label{tab:dataset_stat}}
\resizebox{\columnwidth}{!}{%
\begin{tabular}{|cccc|}
\hline
\multicolumn{1}{|c|}{\textbf{Characteristic}} & \multicolumn{1}{c|}{\textbf{Training}} & \multicolumn{1}{c|}{\textbf{Internal Testing}} & \textbf{External Testing} \\ \hline
\multicolumn{4}{|l|}{Overall characteristic (n = total number)}                                                  \\ \hline
\multicolumn{1}{|c|}{Number of Patients}        & \multicolumn{1}{c|}{7,623}  & \multicolumn{1}{c|}{310}   & 147  \\ \hline
\multicolumn{1}{|c|}{Number of Studies}         & \multicolumn{1}{c|}{7,623}  & \multicolumn{1}{c|}{310}   & 147 \\ \hline
\multicolumn{1}{|c|}{Number of Images}          & \multicolumn{1}{c|}{11,722} & \multicolumn{1}{c|}{1,227} & 220 \\ \hline
\multicolumn{4}{|l|}{View characteristics}                                                                       \\ \hline
\multicolumn{1}{|c|}{MLO View}                  & \multicolumn{1}{c|}{6,409}  & \multicolumn{1}{c|}{287}   & 121 \\ \hline
\multicolumn{1}{|c|}{CC View}                   & \multicolumn{1}{c|}{5,313}  & \multicolumn{1}{c|}{294}   & 99  \\ \hline
\multicolumn{4}{|l|}{Diagnosis metadata}                                                                         \\ \hline
\multicolumn{1}{|c|}{BIRADS 0}                  & \multicolumn{1}{c|}{1,763}  & \multicolumn{1}{c|}{103}   & \multicolumn{1}{c|}{9} \\ \hline
\multicolumn{1}{|c|}{BIRADS 1}                  & \multicolumn{1}{c|}{3,523}  & \multicolumn{1}{c|}{149}   & \multicolumn{1}{c|}{83}  \\ \hline
\multicolumn{1}{|c|}{BIRADS 2}                  & \multicolumn{1}{c|}{339}    & \multicolumn{1}{c|}{15}    & \multicolumn{1}{c|}{129}   \\ \hline
\multicolumn{4}{|l|}{Tissue density metadata}                                                                    \\ \hline
\multicolumn{1}{|c|}{Fatty Density (A)}         & \multicolumn{1}{c|}{577}    & \multicolumn{1}{c|}{18}    & \multicolumn{1}{c|}{18}  \\ \hline
\multicolumn{1}{|c|}{Fibrogranular Density (B)} & \multicolumn{1}{c|}{2,734}  & \multicolumn{1}{c|}{122}   & \multicolumn{1}{c|}{69}  \\ \hline
\multicolumn{1}{|c|}{Heterogeneously Dense (C)} & \multicolumn{1}{c|}{2,647}  & \multicolumn{1}{c|}{95}    & \multicolumn{1}{c|}{100}  \\ \hline
\multicolumn{1}{|c|}{Dense Tissue (D)}          & \multicolumn{1}{c|}{262}    & \multicolumn{1}{c|}{20}    & \multicolumn{1}{c|}{27}  \\ \hline
\multicolumn{4}{|l|}{OOD type}                                                                                   \\ \hline
\multicolumn{1}{|c|}{Implant}                   & \multicolumn{1}{c|}{0}      & \multicolumn{1}{c|}{342}   & 64  \\ \hline
\multicolumn{1}{|c|}{Poor Quality}              & \multicolumn{1}{c|}{0}      & \multicolumn{1}{c|}{200}   & 40  \\ \hline
\multicolumn{1}{|c|}{Biopsy Clip}               & \multicolumn{1}{c|}{0}      & \multicolumn{1}{c|}{-}     & 39  \\ \hline
\end{tabular}%
}

\end{table}

For both datasets, left and right paired MLO views are identified and stitched together while maintaining the 2:1 aspect ratio. Images are standardized within 0-1 intensity range after min-max normalization. 

\subsection{Baselines}
Given the unsupervised nature of the study (no anomaly training data), we compared the proposed architecture against state-of-the-art unsupervised generative models - (i) encoder-based: CVAE\cite{pol2019anomaly}, CVAD\cite{guo2022cvad}, VQ-VAE\cite{marimont2021anomaly} and (ii) GAN based:  f-AnoGAN~\cite{schlegl2019f}. The baseline models are trained and validated on the exact same internal and external datasets using default hyperparameters and we reported both reconstruction and anomaly detection performance against the proposed hybrid model.    

\subsection{Quantitative Performance}
\begin{figure}[htb!]
\centering
\begin{subfigure}[]{}
    \includegraphics[width=1.0\linewidth]{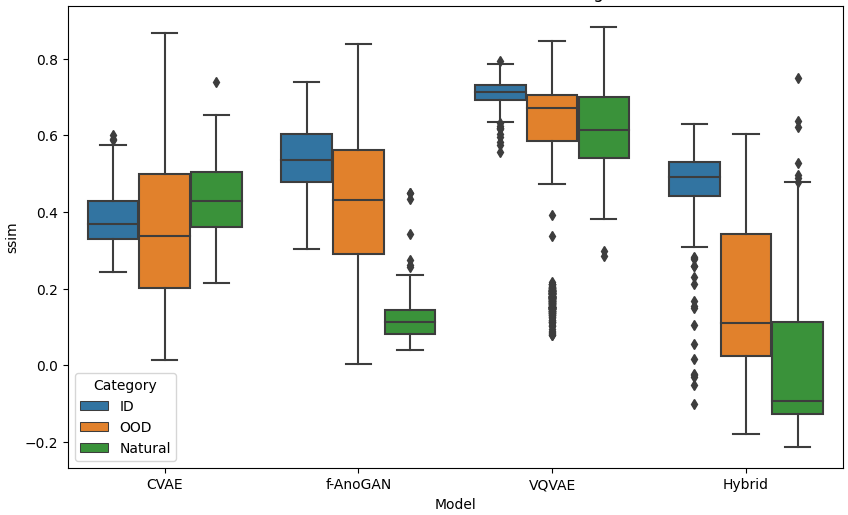}
    \label{fig:first}
\end{subfigure}
\begin{subfigure}[]{}
    \includegraphics[width=1.0\linewidth]{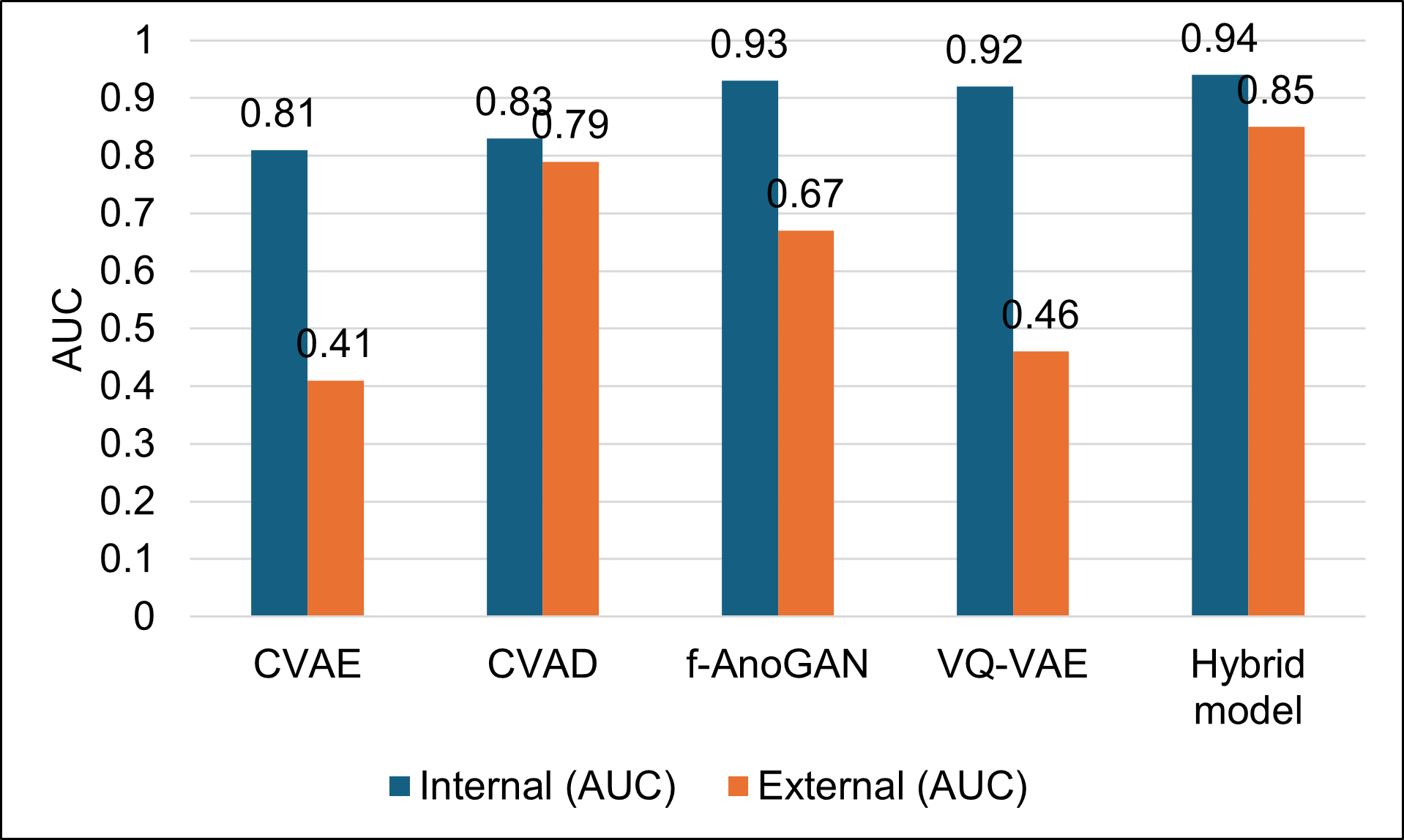}
    \label{fig:second}
\end{subfigure}
\caption{Comparative analysis of various models - (a) SSIM values for different models for ID, ODD and natural image data; (b) OOD classification performance reported in terms of AUC on the internal and external dataset.}
\label{fig:ssim}
\end{figure}

\noindent \emph{Reconstruction Performance}
We measure the SSIM (Structure Similarity Index Measure) score to evaluate the quality of reconstructed image from the original input image. The higher SSIM score ($>0.8$) indicates the reconstructed image is more similar with the original input image. Fig.\ref{fig:ssim}. presents the SSIM score of test dataset across different models. For each model, the test dataset is divided into three categories for SSIM evaluation - ID is the in-distribution mammograms, OOD includes out-of-distribution mammograms cases, such as implant, implant rapture, poor quality, and noises added to mammograms. Natural category include the images from ImageNet which are not mammograms. 

For anomaly detection using generative models, a high SSIM score for the ODD mammograms and natural images suggests that the model cannot differentiate between ID and ODD when reconstructing the images. A good anomaly detection model should have a much higher SSIM score for the ID category than the other two categories, which shows a clear separation between ID data and others. 
Following that notion, VQVAE perform worsen ($>0.7$ average SSIM score for ODD) than f-AnoGAN ($>0.55$ average SSIM score for ODD) and hybrid models ($<0.01$ average SSIM score for ODD) as shown in Fig.\ref{fig:ssim}. f-AnoGAN and hybrid model follow the pattern of obtaining ID SSIM score much higher than OOD and natural categories. Having a more distinctive SSIM score between ID and ODD leads to higher reconstruction loss for OOD and natural samples. Although, as described in Section \ref{sec3}, the model's reconstruction ability alone is not sufficient to use as anomaly score since SSIM score for ID outliers significantly overlaps with ODDs - even for the hybrid model. \\

\noindent \emph{Anomaly Detection Performance} 
Figure \ref{fig:ssim}.b. and Table \ref{tab:anoEvalauc} shows accuracy and AUC for anomaly detection of comparative baselines and the proposed hybrid model with discriminator branch using gradient reversal. Even though encoder-based models obtained high average accuracy (CVAE - 0.81, VQ-VAE - 0.86 AUROC)  for the internal data, there is a significant drop on the external data (CVAE - 0.4, VQ-VAE - 0.46 AUROC). While CVAD - encoder with discriminator branch obtained 0.83 AUROC on internal and 0.73 AUROC on the external data, f-AnoGAN obtained 0.93 and 0.67 AUROC on internal and external data respectively. Proposed hybrid model -- HAND outperformed all the baselines and obtained 0.93 and 0.85 AUROC on internal and external data. Although there is a minor drop in the performance on the external data, HAND is able to achieve higher performance for all the anomaly sub-types of the external data - implant (0.9), biopsy clip (0.89) and poor quality (0.77). All the baselines along with the proposed model struggle to differentiate the poor quality images from external ID data which could be due to the variation between internal and external data, and heterogeneous and unknown nature of the imaging artifacts.

\begin{table}[]
\caption{Comparative overall and sub-types performance for anomaly detection on both internal and external datasets}\label{tab:anoEvalauc}
\resizebox{\columnwidth}{!}{%
\begin{tabular}{|cccccc|}
\hline
\multicolumn{3}{|c|}{\textbf{Internal Dataset (RSNA)}} & \multicolumn{3}{c|}{\textbf{External Dataset (Mayo)}} \\
\hline
\multicolumn{1}{|c|}{\textit{Characteristic}}  & \multicolumn{1}{c|}{\textit{Accuracy}} & \multicolumn{1}{c|}{\textit{AUROC}}
& \multicolumn{1}{c|}{\textit{Characteristic}} & \multicolumn{1}{c|}{\textit{Accuracy}} & \multicolumn{1}{c|}{\textit{AUROC}} \\ \hline
\multicolumn{6}{|c|}{\textbf{Model: CVAE}} \\ \hline
\multicolumn{1}{|c|}{Overall} & \multicolumn{1}{c|}{0.753} & \multicolumn{1}{c|}{0.814} & \multicolumn{1}{c|}{Overall} & \multicolumn{1}{c|}{0.409} & \multicolumn{1}{c|}{0.407}\\ \hline
\multicolumn{1}{|c|}{Normal (ID)}  & \multicolumn{1}{c|}{1.000} & \multicolumn{1}{c|}{-} & \multicolumn{1}{c|}{Normal (ID)} & \multicolumn{1}{c|}{1.000} & \multicolumn{1}{c|}{-} \\ \hline
\multicolumn{1}{|c|}{Implant}  & \multicolumn{1}{c|}{0.840} & \multicolumn{1}{c|}{0.929} & \multicolumn{1}{c|}{Implant} &  \multicolumn{1}{c|}{0.440} & \multicolumn{1}{c|}{0.409}\\ \hline
\multicolumn{1}{|c|}{Noise}  &  \multicolumn{1}{c|}{0.957} & \multicolumn{1}{c|}{0.990} & \multicolumn{1}{c|}{Biopsy Clip} & \multicolumn{1}{c|}{0.379} & \multicolumn{1}{c|}{0.381}\\ \hline
\multicolumn{1}{|c|}{Poor Quality} & \multicolumn{1}{c|}{0.514} & \multicolumn{1}{c|}{0.281} & \multicolumn{1}{c|}{Poor Quality} & \multicolumn{1}{c|}{0.521} & \multicolumn{1}{c|}{0.429}\\ \hline
\multicolumn{1}{|c|}{Natural Image} & \multicolumn{1}{c|}{0.968} & \multicolumn{1}{c|}{0.987} & \multicolumn{1}{c|}{-} & \multicolumn{1}{c|}{-} & \multicolumn{1}{c|}{-} \\ \hline

\multicolumn{6}{|c|}{\textbf{Model: CVAD}} \\ \hline
\multicolumn{1}{|c|}{Overall}  & \multicolumn{1}{c|}{0.833} & \multicolumn{1}{c|}{0.826} & \multicolumn{1}{c|}{Overall} & \multicolumn{1}{c|}{0.668} & \multicolumn{1}{c|}{0.728}\\ \hline
\multicolumn{1}{|c|}{Normal (ID)}  & \multicolumn{1}{c|}{1.000} & \multicolumn{1}{c|}{-} & \multicolumn{1}{c|}{Normal (ID)} & \multicolumn{1}{c|}{1.000} & \multicolumn{1}{c|}{-} \\ \hline
\multicolumn{1}{|c|}{Implant}  & \multicolumn{1}{c|}{0.756} & \multicolumn{1}{c|}{0.780} & \multicolumn{1}{c|}{Implant} & \multicolumn{1}{c|}{0.723} & \multicolumn{1}{c|}{0.711} \\ \hline
\multicolumn{1}{|c|}{Noise} & \multicolumn{1}{c|}{0.857} & \multicolumn{1}{c|}{0.885} & \multicolumn{1}{c|}{Biopsy Clip}  & \multicolumn{1}{c|}{0.690} & \multicolumn{1}{c|}{0.747}\\ \hline
\multicolumn{1}{|c|}{Poor Quality}  & \multicolumn{1}{c|}{0.843} & \multicolumn{1}{c|}{0.884} & \multicolumn{1}{c|}{Poor Quality} & \multicolumn{1}{c|}{0.598} & {0.643}\\ \hline
\multicolumn{1}{|c|}{Natural Image} & \multicolumn{1}{c|}{0.828} & \multicolumn{1}{c|}{0.864} & \multicolumn{1}{c|}{-} & \multicolumn{1}{c|}{-} & \multicolumn{1}{c|}{-} \\ \hline

\multicolumn{6}{|c|}{\textbf{Model: f-AnoGAN}} \\ \hline
\multicolumn{1}{|c|}{Overall} & \multicolumn{1}{c|}{0.8826} & \multicolumn{1}{c|}{0.929} & \multicolumn{1}{c|}{Overall}  & \multicolumn{1}{c|}{0.6227} & \multicolumn{1}{c|}{0.672}\\ \hline
\multicolumn{1}{|c|}{Normal (ID)}  & \multicolumn{1}{c|}{1.000} & \multicolumn{1}{c|}{-} & \multicolumn{1}{c|}{Normal (ID)} & \multicolumn{1}{c|}{1.000} & \multicolumn{1}{c|}{-} \\ \hline
\multicolumn{1}{|c|}{Implant}  & \multicolumn{1}{c|}{0.882} & \multicolumn{1}{c|}{0.952} & \multicolumn{1}{c|}{Implant} & {0.631} & \multicolumn{1}{c|}{0.719}\\ \hline
\multicolumn{1}{|c|}{Noise}  & \multicolumn{1}{c|}{0.967} & \multicolumn{1}{c|}{0.993} & \multicolumn{1}{c|}{Biopsy Clip} & \multicolumn{1}{c|}{0.690} & \multicolumn{1}{c|}{0.644}\\ \hline
\multicolumn{1}{|c|}{Poor Quality} & \multicolumn{1}{c|}{0.841} & \multicolumn{1}{c|}{0.761} & \multicolumn{1}{c|}{Poor Quality} & \multicolumn{1}{c|}{0.581} & {0.624}\\ \hline
\multicolumn{1}{|c|}{Natural Image}  & \multicolumn{1}{c|}{0.980} & \multicolumn{1}{c|}{0.999} & \multicolumn{1}{c|}{-} & \multicolumn{1}{c|}{-} & \multicolumn{1}{c|}{-} \\ \hline

\multicolumn{6}{|c|}{\textbf{Model: VQ-VAE}} \\ \hline
\multicolumn{1}{|c|}{Overall}  & \multicolumn{1}{c|}{0.861} & \multicolumn{1}{c|}{0.922} & \multicolumn{1}{c|}{Overall}  & \multicolumn{1}{c|}{0.4773} & \multicolumn{1}{c|}{0.463}\\ \hline
\multicolumn{1}{|c|}{Normal (ID)}  & \multicolumn{1}{c|}{1.000} & \multicolumn{1}{c|}{-} & \multicolumn{1}{c|}{Normal (ID)} & \multicolumn{1}{c|}{1.000} & \multicolumn{1}{c|}{-} \\ \hline
\multicolumn{1}{|c|}{Implant}  & \multicolumn{1}{c|}{0.830} & \multicolumn{1}{c|}{0.862} & \multicolumn{1}{c|}{Implant} & \multicolumn{1}{c|}{0.624} & {0.486}\\ \hline
\multicolumn{1}{|c|}{Noise}  & \multicolumn{1}{c|}{0.959} & {0.991} & \multicolumn{1}{c|}{Biopsy Clip} & \multicolumn{1}{c|}{0.4741} & \multicolumn{1}{c|}{0.439}\\ \hline
\multicolumn{1}{|c|}{Poor Quality} & \multicolumn{1}{c|}{0.811} & \multicolumn{1}{c|}{0.870} & \multicolumn{1}{c|}{Poor Quality} & \multicolumn{1}{c|}{0.607} & \multicolumn{1}{c|}{0.418}\\ \hline
\multicolumn{1}{|c|}{Natural Image}  & \multicolumn{1}{c|}{0.978} & \multicolumn{1}{c|}{0.996} &  \multicolumn{1}{c|}{-} & \multicolumn{1}{c|}{-} & \multicolumn{1}{c|}{-} \\ \hline

\multicolumn{6}{|c|}{\textbf{Model: HAND}} \\ \hline
\multicolumn{1}{|c|}{Overall}  & \multicolumn{1}{c|}{0.890} & \multicolumn{1}{c|}{0.937} & \multicolumn{1}{c|}{Overall} & \multicolumn{1}{c|}{0.782} & \multicolumn{1}{c|}{0.850}\\ \hline
\multicolumn{1}{|c|}{Normal (ID)} & \multicolumn{1}{c|}{1.000} & \multicolumn{1}{c|}{-} & \multicolumn{1}{c|}{Normal (ID)} & \multicolumn{1}{c|}{1.000} & \multicolumn{1}{c|}{-} \\ \hline
\multicolumn{1}{|c|}{Implant} & \multicolumn{1}{c|}{0.837} & \multicolumn{1}{c|}{0.910} & \multicolumn{1}{c|}{Implant} & \multicolumn{1}{c|}{0.830} & \multicolumn{1}{c|}{0.899}\\ \hline
\multicolumn{1}{|c|}{Noise} & \multicolumn{1}{c|}{0.900} & \multicolumn{1}{c|}{0.935} & \multicolumn{1}{c|}{Biopsy Clip} & \multicolumn{1}{c|}{0.871} & \multicolumn{1}{c|}{0.887}\\ \hline
\multicolumn{1}{|c|}{Poor Quality}  & \multicolumn{1}{c|}{0.942} & \multicolumn{1}{c|}{0.966} & \multicolumn{1}{c|}{Poor Quality} & \multicolumn{1}{c|}{0.7607} & \multicolumn{1}{c|}{0.770}\\ \hline
\multicolumn{1}{|c|}{Natural Image} & \multicolumn{1}{c|}{0.926} & \multicolumn{1}{c|}{0.986} &  \multicolumn{1}{c|}{-} & \multicolumn{1}{c|}{-} & \multicolumn{1}{c|}{-} \\ \hline

\end{tabular}
}
% \footnotetext{This table is the evaluation results for anomaly detection using different models.}
% \footnotetext[1]{}
\end{table}

\subsection{Qualitative Performance}
Table \ref{tab:recon_in} and \ref{tab:recon_ex} demonstrate the reconstruction quality of different models on the same input internal and external datasets. We visualize both ID and various categories of OODs. As we can observe that encoder-based model produced similar reconstruction of OOD as of the ID data, even for the natural images. However, the reconstruction quality for ODD images are worst than ID for the f-AnoGAN and hybrid model as learning on the ID distribution margin, the models aim to project a object shape close to two-sided breast MLOs even for the natural images which resulted higher reconstruction loss for OOD. Reconstruction quality for f-AnoGAN was consistently poor both ID and ODD data on the external Mayo dataset which also reflects on the poor anomaly detection performance.
\begin{table}
\caption{Internal reconstruction example across different models}\label{tab:recon_in}
\resizebox{\columnwidth}{!}{%
  %\begin{adjustbox}
  \begin{tabular}{|c|c|c|c|c|c|}
    \hline
   \textbf{Category} & \textbf{Input} & \textbf{CVAE}\textbf{/D} & \textbf{f-AnoGAN} & \textbf{VQVAE} & \textbf{HAND}  \\
    \hline
    Normal (ID) & 
    \includegraphics[width=0.12\linewidth]{images/recon_flag/internal/0_id.png} & 
    \includegraphics[width=0.12\linewidth]{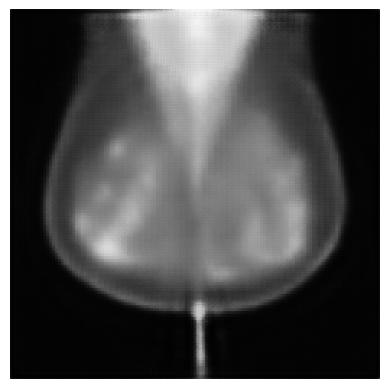} & 
    \includegraphics[width=0.12\linewidth]{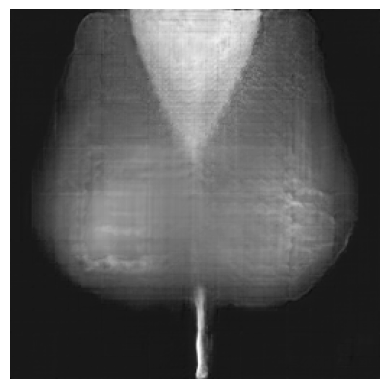} & 
    \includegraphics[width=0.12\linewidth]{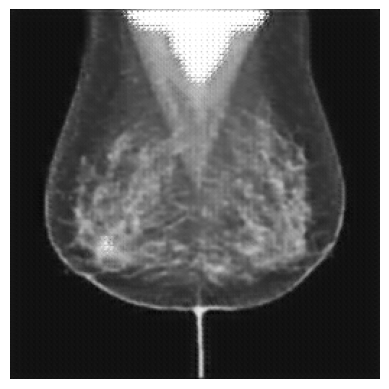} & 
    \includegraphics[width=0.12\linewidth]{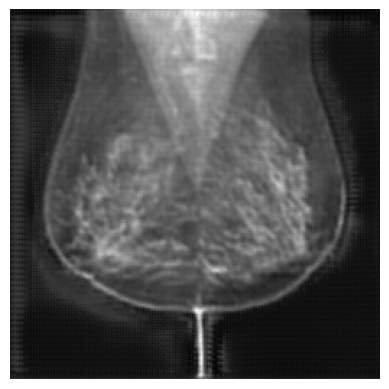} \\
    \hline
    Implant & 
    \includegraphics[width=0.12\linewidth]{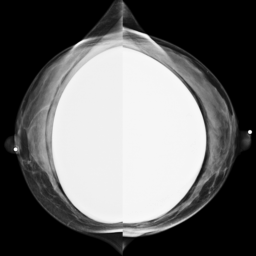} & 
    \includegraphics[width=0.12\linewidth]{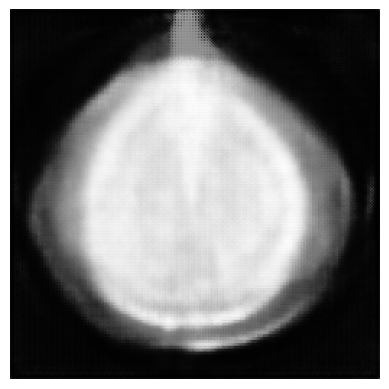} & 
    \includegraphics[width=0.12\linewidth]{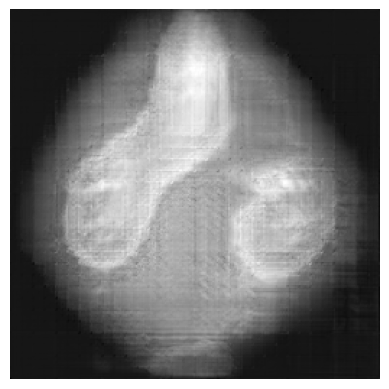} & 
    \includegraphics[width=0.12\linewidth]{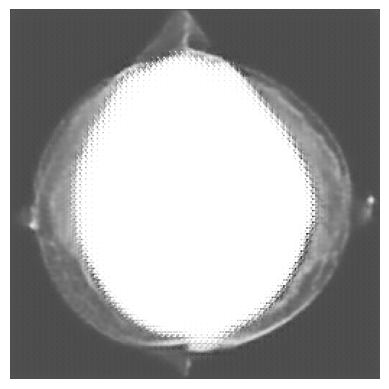} & 
    \includegraphics[width=0.12\linewidth]{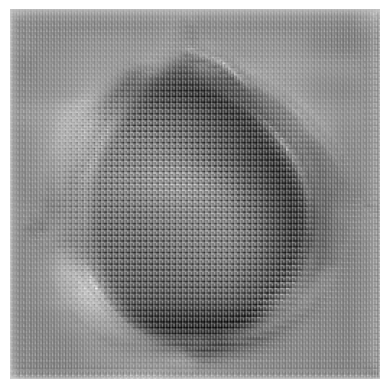} \\
    \hline
    Salt and Pepper & 
    \includegraphics[width=0.12\linewidth]{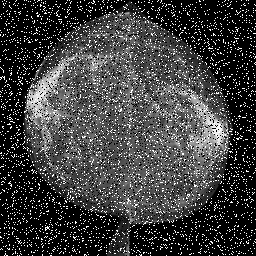} & 
    \includegraphics[width=0.12\linewidth]{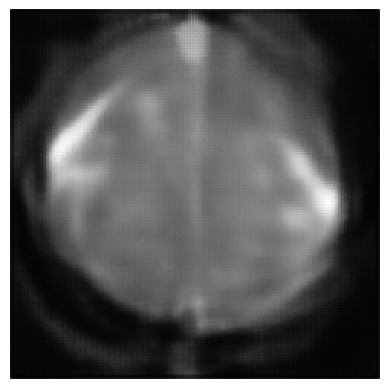} & 
    \includegraphics[width=0.12\linewidth]{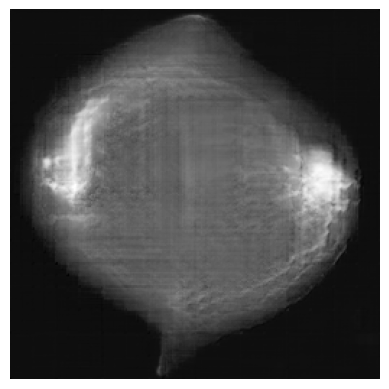} & 
    \includegraphics[width=0.12\linewidth]{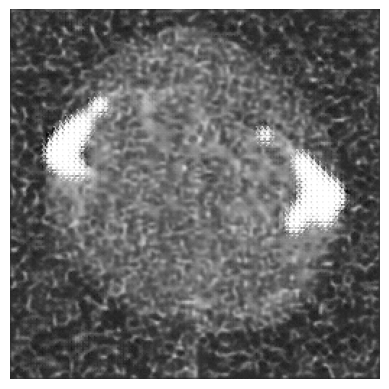} & 
    \includegraphics[width=0.12\linewidth]{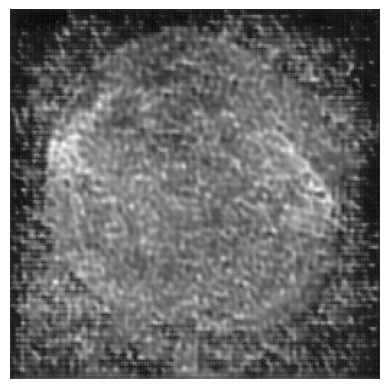} \\
    \hline
    Distortion & 
    \includegraphics[width=0.12\linewidth]{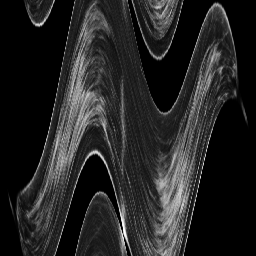} & 
    \includegraphics[width=0.12\linewidth]{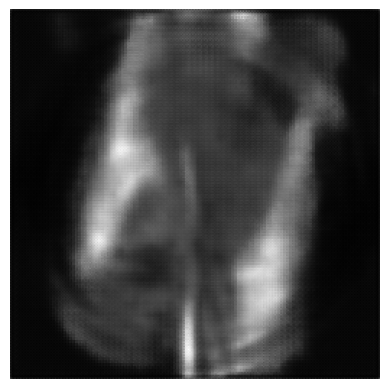} & 
    \includegraphics[width=0.12\linewidth]{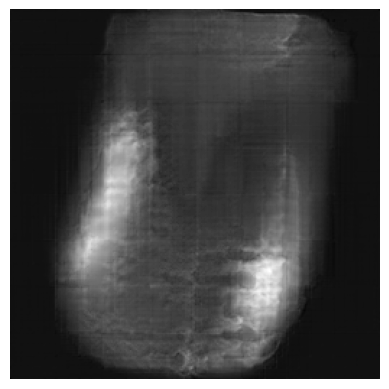} & 
    \includegraphics[width=0.12\linewidth]{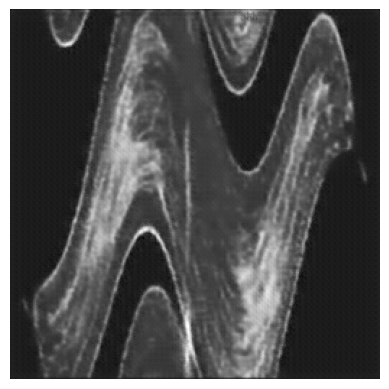} & 
    \includegraphics[width=0.12\linewidth]{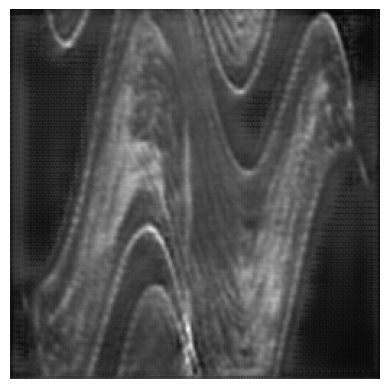} \\
    \hline
    Gaussian & 
    \includegraphics[width=0.12\linewidth]{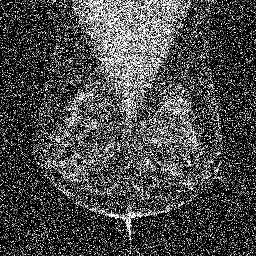} & 
    \includegraphics[width=0.12\linewidth]{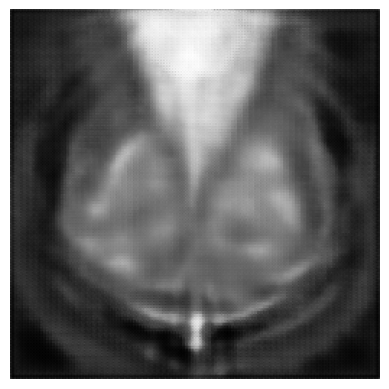} & 
    \includegraphics[width=0.12\linewidth]{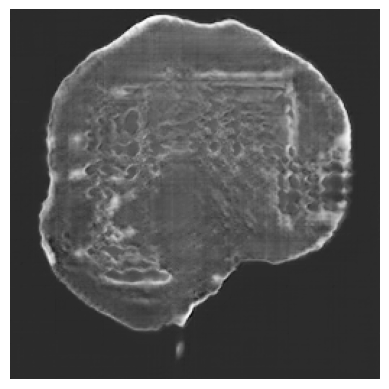} & 
    \includegraphics[width=0.12\linewidth]{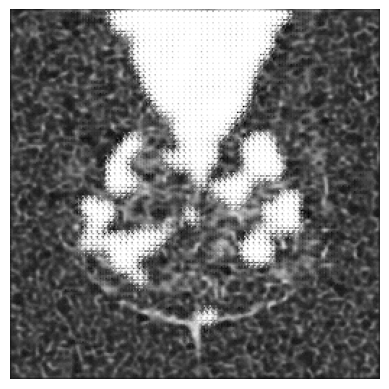} & 
    \includegraphics[width=0.12\linewidth]{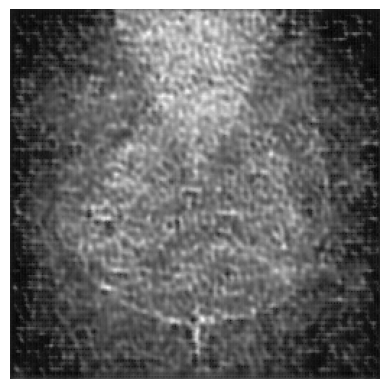} \\
    \hline
    Poor Quality & 
    \includegraphics[width=0.12\linewidth]{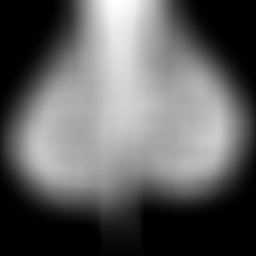} & 
    \includegraphics[width=0.12\linewidth]{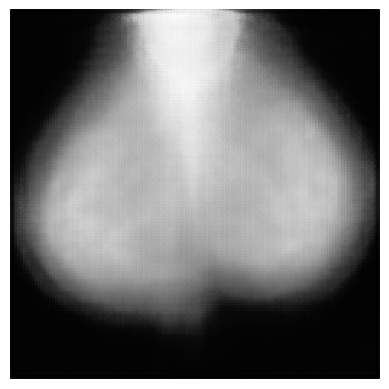} & 
    \includegraphics[width=0.12\linewidth]{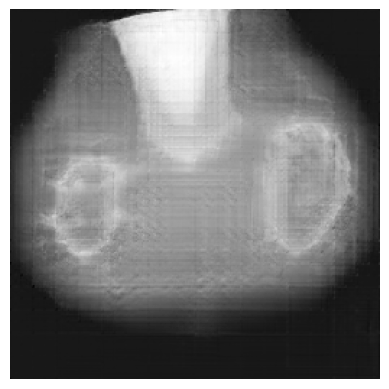} & 
    \includegraphics[width=0.12\linewidth]{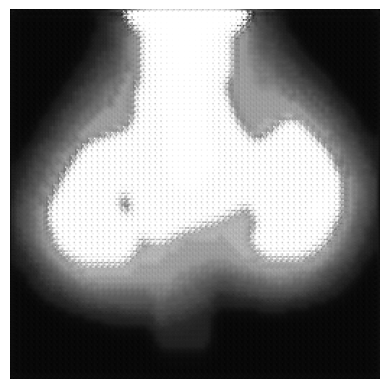} & 
    \includegraphics[width=0.12\linewidth]{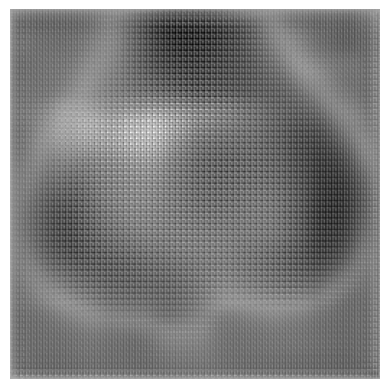} \\
    \hline
    Natural & 
    \includegraphics[width=0.12\linewidth]{images/recon_flag/internal/6_natural.png} & 
    \includegraphics[width=0.12\linewidth]{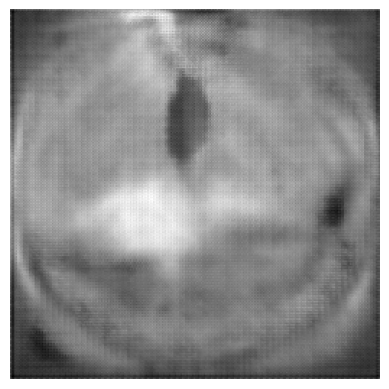} & 
    \includegraphics[width=0.12\linewidth]{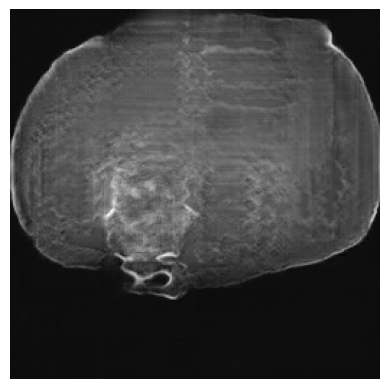} & 
    \includegraphics[width=0.12\linewidth]{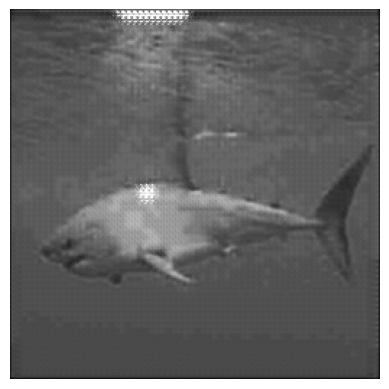} & 
    \includegraphics[width=0.12\linewidth]{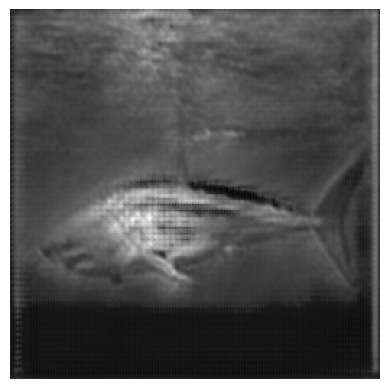} \\
    \hline
  \end{tabular}
  }
  %\end{adjustbox}
  
\end{table}

% \begin{landscape}
\begin{table}[htb!]
\caption{External reconstruction samples across different models} \label{tab:recon_ex}
\resizebox{\columnwidth}{!}{%
  \centering
  %\begin{adjustbox}
  \begin{tabular}{|c|c|c|c|c|c|}
    \hline
     \textbf{Category} & \textbf{Input} & \textbf{CVAE}\textbf{/D} & \textbf{f-AnoGAN} & \textbf{VQVAE} & \textbf{HAND}  \\
    \hline
    Normal (ID) & 
    \includegraphics[width=0.12\linewidth]{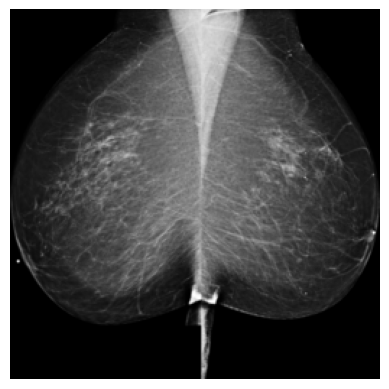} & 
    \includegraphics[width=0.12\linewidth]{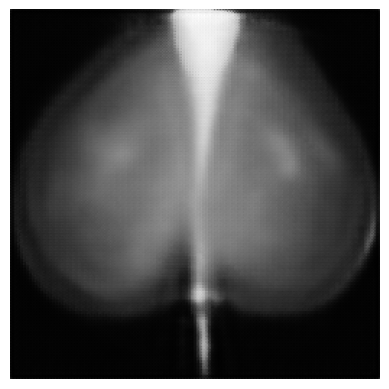} & 
    \includegraphics[width=0.12\linewidth]{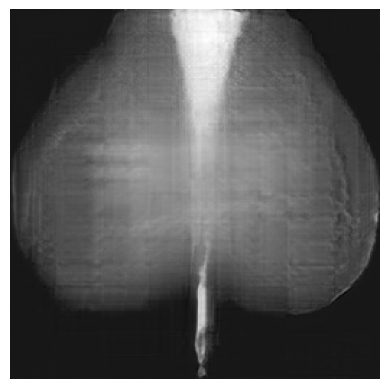} & 
    \includegraphics[width=0.12\linewidth]{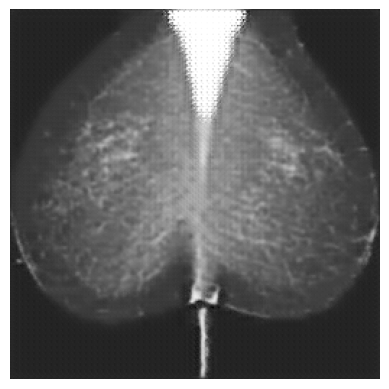} & 
    \includegraphics[width=0.12\linewidth]{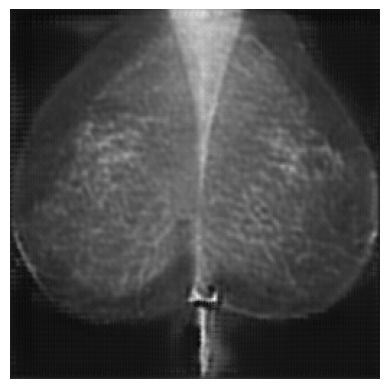} \\
    \hline
    Implant & 
    \includegraphics[width=0.12\linewidth]{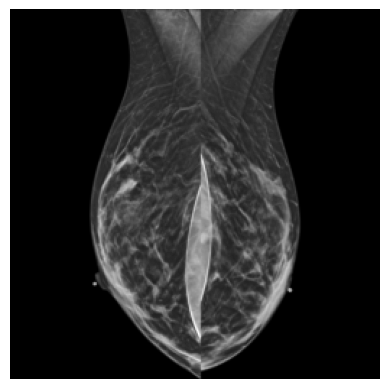} & 
    \includegraphics[width=0.12\linewidth]{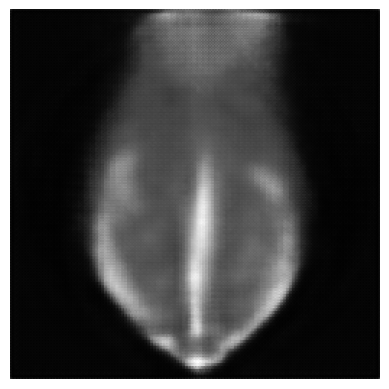} & 
    \includegraphics[width=0.12\linewidth]{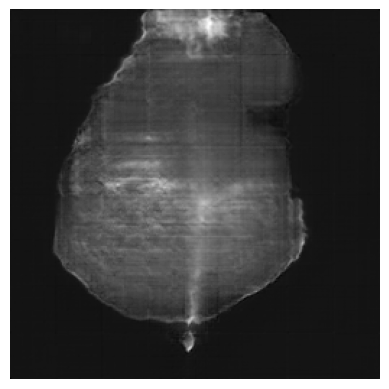} & 
    \includegraphics[width=0.12\linewidth]{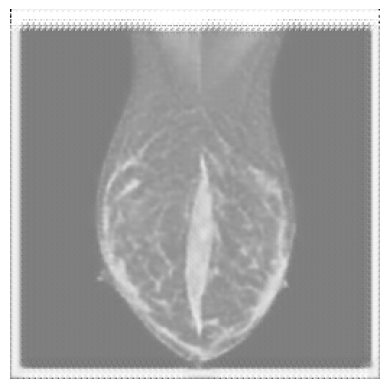} & 
    \includegraphics[width=0.12\linewidth]{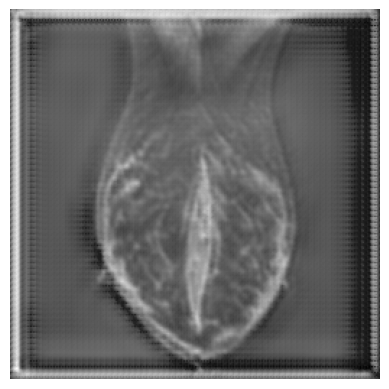} \\
    \hline
    Implant rapture & 
    \includegraphics[width=0.12\linewidth]{images/recon_flag/external/mayo_implantrapture.png} & 
    \includegraphics[width=0.12\linewidth]{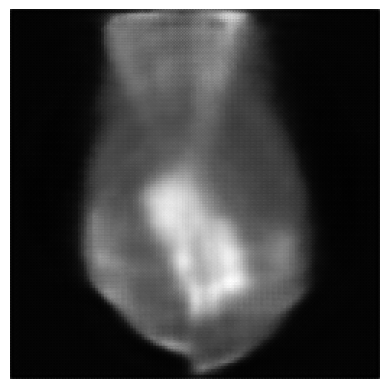} & 
    \includegraphics[width=0.12\linewidth]{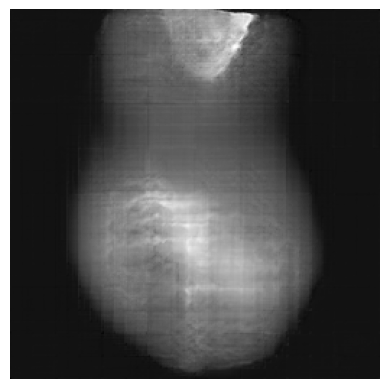} & 
    \includegraphics[width=0.12\linewidth]{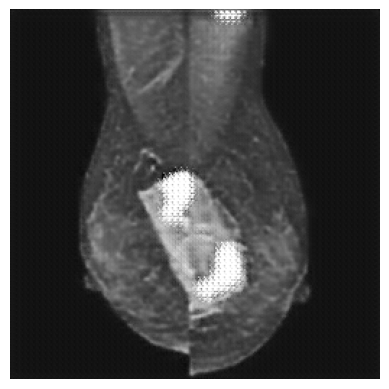} & 
    \includegraphics[width=0.12\linewidth]{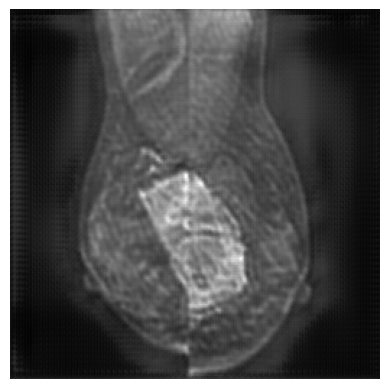} \\
    \hline
    Biopsy Clip & 
    \includegraphics[width=0.12\linewidth]{images/recon_flag/external/mayo_bc.png} & 
    \includegraphics[width=0.12\linewidth]{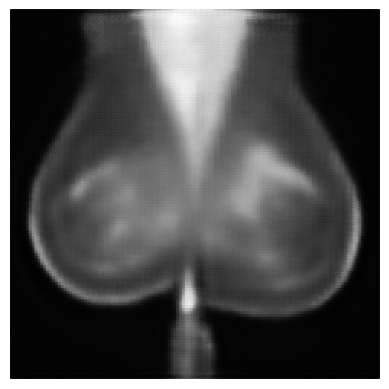} & 
    \includegraphics[width=0.12\linewidth]{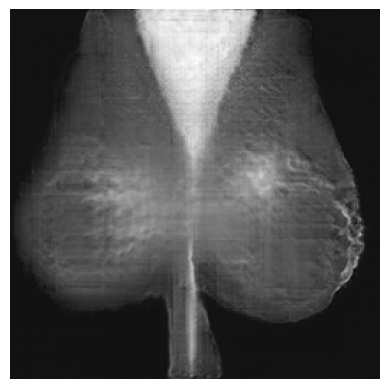} & 
    \includegraphics[width=0.12\linewidth]{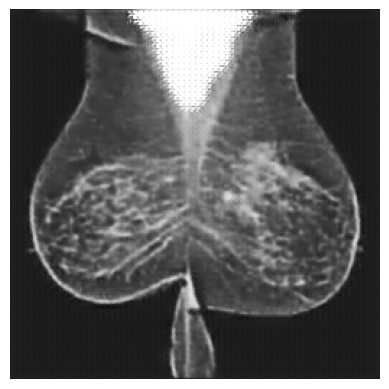} & 
    \includegraphics[width=0.12\linewidth]{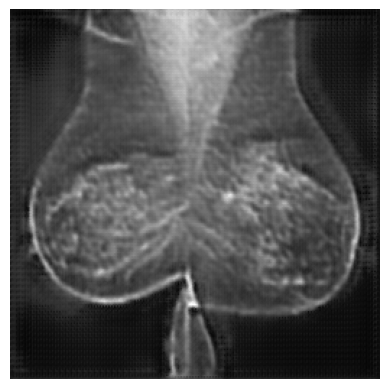} \\
    \hline
    Poor Quality & 
    \includegraphics[width=0.12\linewidth]{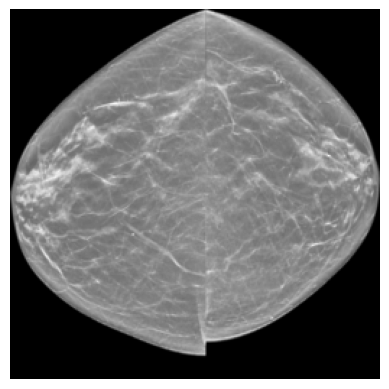} & 
    \includegraphics[width=0.12\linewidth]{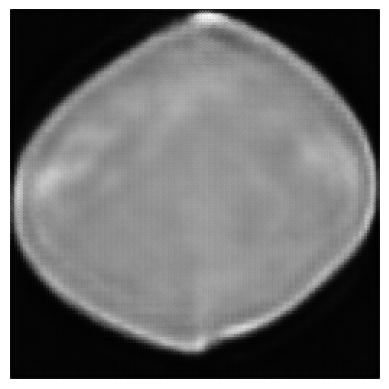} & 
    \includegraphics[width=0.12\linewidth]{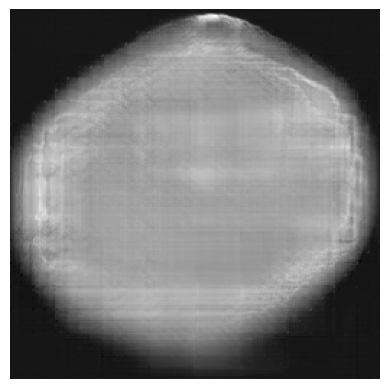} & 
    \includegraphics[width=0.12\linewidth]{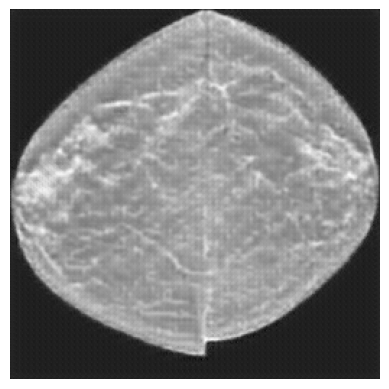} & 
    \includegraphics[width=0.12\linewidth]{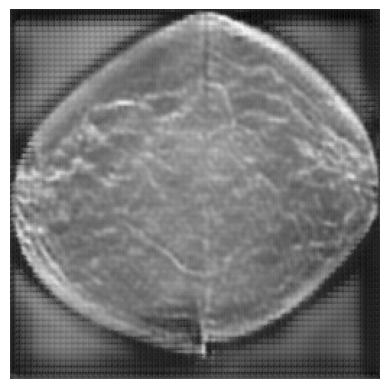} \\
    \hline
  \end{tabular}
  %\end{adjustbox}
  }
\end{table}
% \end{landscape}

\subsection{Ablation Study}
We performed a through ablation study to understand the  contribution of core design components to the overall system performance - (i) model training discriminator loss in addition to reconstruction, (ii) gradient reversal for penalizing the model for learning OOD reconstruction. First, we evaluated the base hybrid TransBTS architecture~\cite{li2022transbtsv2} with image reconstruction mean square error loss. Second, we reported the performance of base architecture trained with the synthetic ODD samples and the discriminator branch. Finally, we report the performance of the proposed architecture with gradient reversal. 

Table \ref{tab:ablationEval} reported the quantitative performance for the ablation settings and Table \ref{tab:recon_in_ablation} shows reconstruction sample of original hybrid model, hybrid model add discriminator, and hybrid model add discriminator and gradient reversal. Base hybrid model~\cite{li2022transbtsv2} achieved similar performance as of the encoder-based model on the internal data (0.87 AUC) and the performance dropped significantly for the external cases (0.56 AUC). Adding the discriminator branch, the performance on both internal and external data improved upto 0.93 and 0.73 respectively; however the performance on small anomalies (e.g. biopsy clip) was still low - 0.63 AUC. Overall performance improved with negative gradient on both internal (0.93 AUC) and external (0.85 AUC) dataset, even for the small anomalies (e.g. biopsy clip 0.89 AUC). Reconstruction example also highlight the impact of discriminator with negative gradient, particularly for implant and natural images. 

\begin{table}[]
\caption{Evaluation of anomaly detection performance for ablation study}\label{tab:ablationEval}
\resizebox{\columnwidth}{!}{%
\begin{tabular}{|cccccc|}
\hline
\multicolumn{3}{|c|}{\textbf{Internal Dataset (RSNA)}} & \multicolumn{3}{c|}{\textbf{External Dataset (Mayo)}} \\
\hline
\multicolumn{1}{|c|}{\textit{Characteristic}}  & \multicolumn{1}{c|}{\textit{Accuracy}} & \multicolumn{1}{c|}{\textit{AUROC}}
& \multicolumn{1}{c|}{\textit{Characteristic}} & \multicolumn{1}{c|}{\textit{Accuracy}} & \multicolumn{1}{c|}{\textit{AUROC}} \\ \hline
\multicolumn{6}{|c|}{\textbf{Model: Hybrid Baseline}} \\ \hline
\multicolumn{1}{|c|}{Overall}  & \multicolumn{1}{c|}{0.800} & \multicolumn{1}{c|}{0.873} & \multicolumn{1}{c|}{Overall} & \multicolumn{1}{c|}{0.596} & \multicolumn{1}{c|}{0.562}\\ \hline
\multicolumn{1}{|c|}{Normal (ID)}  & \multicolumn{1}{c|}{1.000} & \multicolumn{1}{c|}{-} & \multicolumn{1}{c|}{Normal (ID)} & \multicolumn{1}{c|}{1.000} & \multicolumn{1}{c|}{-}\\ \hline
\multicolumn{1}{|c|}{Implant} & \multicolumn{1}{c|}{0.733} & \multicolumn{1}{c|}{0.806} & \multicolumn{1}{c|}{Implant} & \multicolumn{1}{c|}{0.582} & \multicolumn{1}{c|}{0.561}\\ \hline
\multicolumn{1}{|c|}{Noise}  &  \multicolumn{1}{c|}{0.902} & \multicolumn{1}{c|}{0.967} & \multicolumn{1}{c|}{Biopsy Clip} & \multicolumn{1}{c|}{0.621} & \multicolumn{1}{c|}{0.493}\\ \hline
\multicolumn{1}{|c|}{Poor Quality} & \multicolumn{1}{c|}{0.795} & \multicolumn{1}{c|}{0.820} & \multicolumn{1}{c|}{Poor Quality} & \multicolumn{1}{c|}{0.564} & \multicolumn{1}{c|}{0.556}\\ \hline
\multicolumn{1}{|c|}{Natural Image}  & \multicolumn{1}{c|}{0.970} & \multicolumn{1}{c|}{0.994} & \multicolumn{1}{c|}{-} & \multicolumn{1}{c|}{-} & \multicolumn{1}{c|}{-} \\ \hline

\multicolumn{6}{|c|}{\textbf{Model: Hybrid with Discriminator}} \\ \hline
\multicolumn{1}{|c|}{Overall} & \multicolumn{1}{c|}{0.902} & \multicolumn{1}{c|}{0.934} & \multicolumn{1}{c|}{Overall} & \multicolumn{1}{c|}{0.695} & \multicolumn{1}{c|}{0.730}\\ \hline
\multicolumn{1}{|c|}{Normal (ID)}  & \multicolumn{1}{c|}{1.000} & \multicolumn{1}{c|}{-} & \multicolumn{1}{c|}{Normal (ID)} & \multicolumn{1}{c|}{1.000} & \multicolumn{1}{c|}{-} \\ \hline
\multicolumn{1}{|c|}{Implant}  & \multicolumn{1}{c|}{0.843} & \multicolumn{1}{c|}{0.898} & \multicolumn{1}{c|}{Implant} & \multicolumn{1}{c|}{0.809} & \multicolumn{1}{c|}{0.756}\\ \hline
\multicolumn{1}{|c|}{Noise} & \multicolumn{1}{c|}{0.907} & \multicolumn{1}{c|}{0.962} & \multicolumn{1}{c|}{Biopsy Clip} & \multicolumn{1}{c|}{0.690} & \multicolumn{1}{c|}{0.632}\\ \hline
\multicolumn{1}{|c|}{Poor Quality} & \multicolumn{1}{c|}{0.932} & \multicolumn{1}{c|}{0.932} & \multicolumn{1}{c|}{Poor Quality} & \multicolumn{1}{c|}{0.770} & \multicolumn{1}{c|}{0.723}\\ \hline
\multicolumn{1}{|c|}{Natural Image} & \multicolumn{1}{c|}{0.978} & \multicolumn{1}{c|}{0.997} & \multicolumn{1}{c|}{-} & \multicolumn{1}{c|}{-} & \multicolumn{1}{c|}{-} \\ \hline

\multicolumn{6}{|c|}{\textbf{Model: HAND}} \\ \hline
\multicolumn{1}{|c|}{Overall}  & \multicolumn{1}{c|}{0.890} & \multicolumn{1}{c|}{0.937} & \multicolumn{1}{c|}{Overall} & \multicolumn{1}{c|}{0.782} & \multicolumn{1}{c|}{0.850}\\ \hline
\multicolumn{1}{|c|}{Normal (ID)} & \multicolumn{1}{c|}{1.000} & \multicolumn{1}{c|}{-} & \multicolumn{1}{c|}{Normal (ID)} & \multicolumn{1}{c|}{1.000} & \multicolumn{1}{c|}{-} \\ \hline
\multicolumn{1}{|c|}{Implant} & \multicolumn{1}{c|}{0.837} & \multicolumn{1}{c|}{0.910} & \multicolumn{1}{c|}{Implant} & \multicolumn{1}{c|}{0.830} & \multicolumn{1}{c|}{0.899}\\ \hline
\multicolumn{1}{|c|}{Noise}  & \multicolumn{1}{c|}{0.900} & \multicolumn{1}{c|}{0.935} & \multicolumn{1}{c|}{Biopsy Clip} & \multicolumn{1}{c|}{0.871} & \multicolumn{1}{c|}{0.887}\\ \hline
\multicolumn{1}{|c|}{Poor Quality} & \multicolumn{1}{c|}{0.942} & \multicolumn{1}{c|}{0.966} & \multicolumn{1}{c|}{Poor Quality} & \multicolumn{1}{c|}{0.7607} & \multicolumn{1}{c|}{0.770}\\ \hline
\multicolumn{1}{|c|}{Natural Image} & \multicolumn{1}{c|}{0.926} & \multicolumn{1}{c|}{0.986} & \multicolumn{1}{c|}{-} & \multicolumn{1}{c|}{-} & \multicolumn{1}{c|}{-} \\ \hline
%\botrule
\end{tabular}
%\footnotetext{This table is the evaluation results for anomaly detection for ablation study.}
%\footnotetext[1]{}
}
\end{table}

\begin{table}[htb!]
 \caption{Ablation: internal reconstruction samples across different hybrid models}\label{tab:recon_in_ablation}
  \centering
  \resizebox{\columnwidth}{!}{%
  % \begin{adjustbox}{max width=\linewidth, max height=\textheight}
  \begin{tabular}{|c|c|c|c|c|}
    \hline
    & Input & Hybrid & Hybrid w/Dis.& HAND\\
    \hline
    Normal (ID) & 
    \includegraphics[width=0.1\linewidth]{images/recon_flag/internal/0_id.png} & 
    \includegraphics[width=0.1\linewidth]{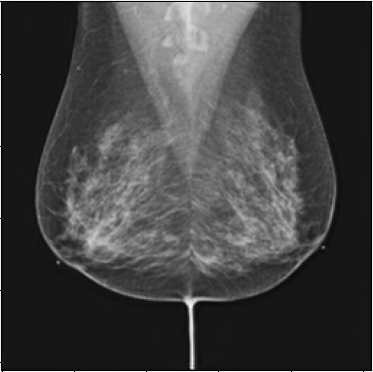} & 
    \includegraphics[width=0.1\linewidth]{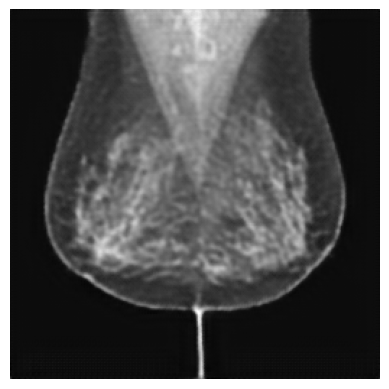} & 
    \includegraphics[width=0.1\linewidth]{images/recon_internal_models/hybrid/internal_id.png} \\ 
    
    \hline
    Implant & 
    \includegraphics[width=0.1\linewidth]{images/recon_flag/internal/1_implant.png} & 
    \includegraphics[width=0.1\linewidth]{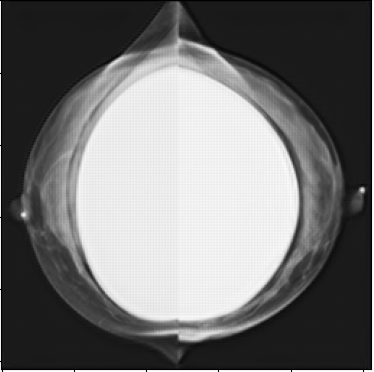} & 
    \includegraphics[width=0.1\linewidth]{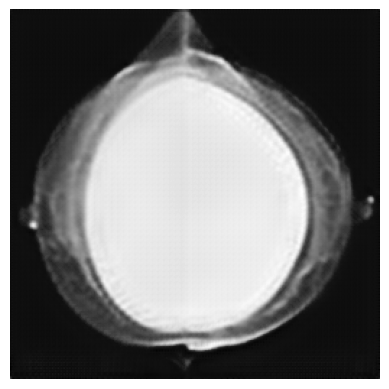} & 
    \includegraphics[width=0.1\linewidth]{images/recon_internal_models/hybrid/internal_implant.png} \\
    
    \hline
    Salt and Pepper & 
    \includegraphics[width=0.1\linewidth]{images/recon_flag/internal/2_sp.png} & 
    \includegraphics[width=0.1\linewidth]{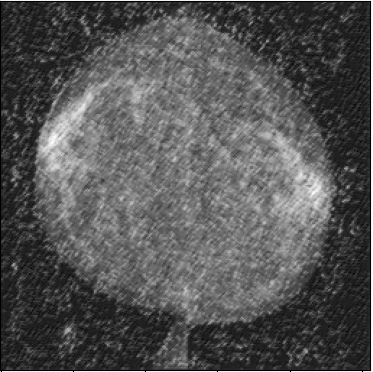} & 
    \includegraphics[width=0.1\linewidth]{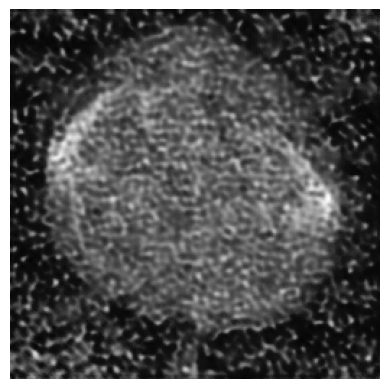} & 
    \includegraphics[width=0.1\linewidth]{images/recon_internal_models/hybrid/internal_sp.png} \\
    
    \hline
    Distortion & 
    \includegraphics[width=0.1\linewidth]{images/recon_flag/internal/3_distort.png} & 
    \includegraphics[width=0.1\linewidth]{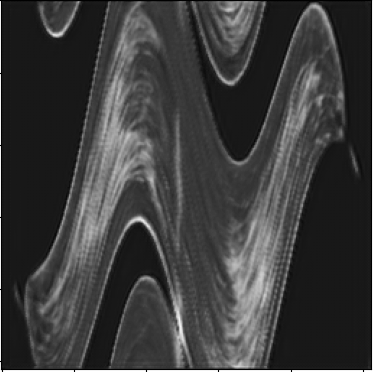} & 
    \includegraphics[width=0.1\linewidth]{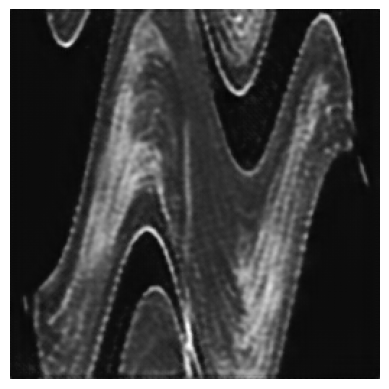} & 
    \includegraphics[width=0.1\linewidth]{images/recon_internal_models/hybrid/internal_dist.png} \\
    
    \hline
    Gaussian & 
    \includegraphics[width=0.1\linewidth]{images/recon_flag/internal/4_gaussian.png} & 
    \includegraphics[width=0.1\linewidth]{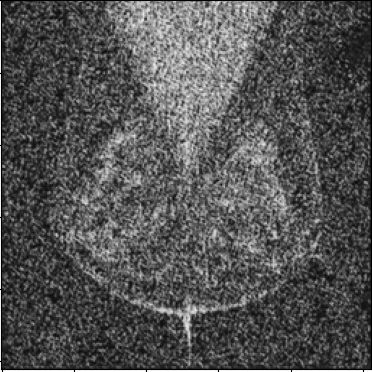} & 
    \includegraphics[width=0.1\linewidth]{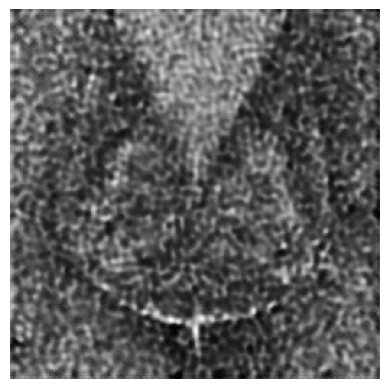} & 
    \includegraphics[width=0.1\linewidth]{images/recon_internal_models/hybrid/internal_gaussian.png} \\
    
    \hline
    Poor Quality & 
    \includegraphics[width=0.1\linewidth]{images/recon_flag/internal/5_poorRes.png} & 
    \includegraphics[width=0.1\linewidth]{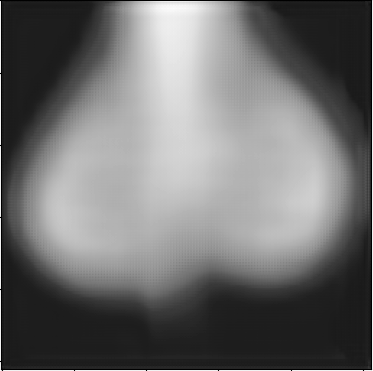} & 
    \includegraphics[width=0.1\linewidth]{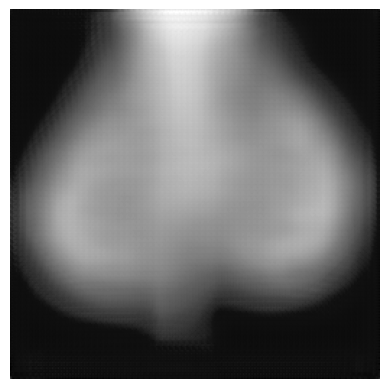} & 
    \includegraphics[width=0.1\linewidth]{images/recon_internal_models/hybrid/internal_poorRes.png} \\
    
    \hline
    Natural & 
    \includegraphics[width=0.1\linewidth]{images/recon_flag/internal/6_natural.png} & 
    \includegraphics[width=0.1\linewidth]{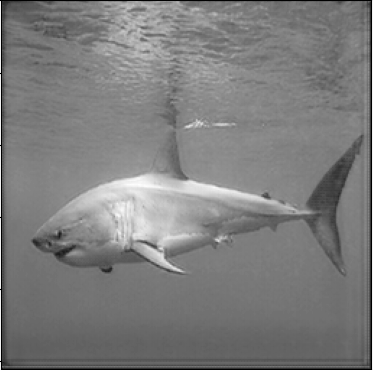} & 
    \includegraphics[width=0.1\linewidth]{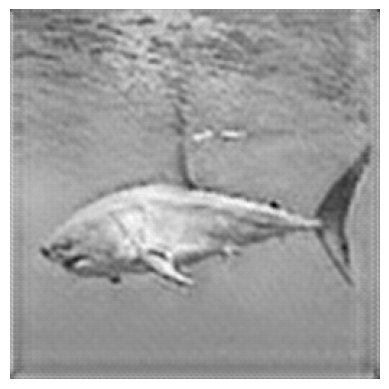} & 
    \includegraphics[width=0.1\linewidth]{images/recon_internal_models/hybrid/internal_natual.png} \\
    \hline
  \end{tabular}
  }
  % \end{adjustbox}
  % \end{adjustbox}
\end{table}

% \begin{landscape}
\begin{table}[h!]
  \caption{Ablation: external reconstruction samples across different hybrid models} \label{tab:recon_ex_ablation}
  \centering
  \resizebox{\columnwidth}{!}{%
  %\begin{adjustbox}
  \begin{tabular}{|c|c|c|c|c|}
    \hline
    & Input & Hybrid & Hybrid w/Dis. & HAND \\
    \hline
    Normal (ID) & 
    \includegraphics[width=0.1\linewidth]{images/recon_flag/external/mayo_null.png} & 
    \includegraphics[width=0.1\linewidth]{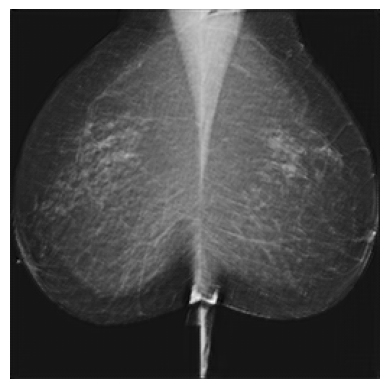} & 
    \includegraphics[width=0.1\linewidth]{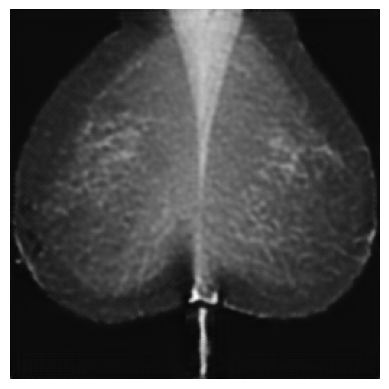} & 
    \includegraphics[width=0.1\linewidth]{images/recon_external_models/hybrid/external_null.png} \\
    \hline
    Implant & 
    \includegraphics[width=0.1\linewidth]{images/recon_flag/external/mayo_implant.png} & 
    \includegraphics[width=0.1\linewidth]{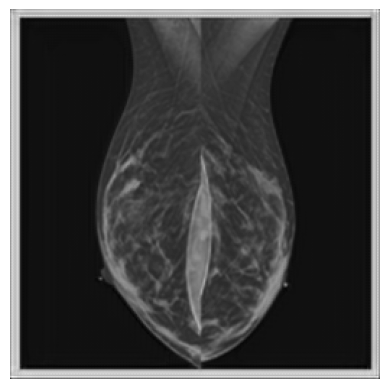} & 
    \includegraphics[width=0.1\linewidth]{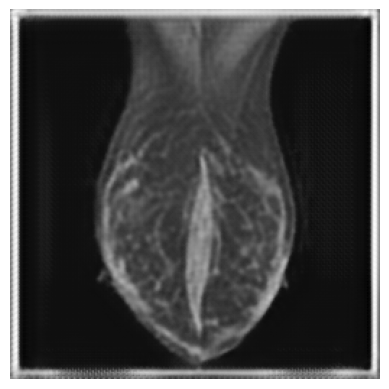} & 
    \includegraphics[width=0.1\linewidth]{images/recon_external_models/hybrid/external_implant.png} \\
    \hline
    Implant rapture & 
    \includegraphics[width=0.1\linewidth]{images/recon_flag/external/mayo_implantrapture.png} & 
    \includegraphics[width=0.1\linewidth]{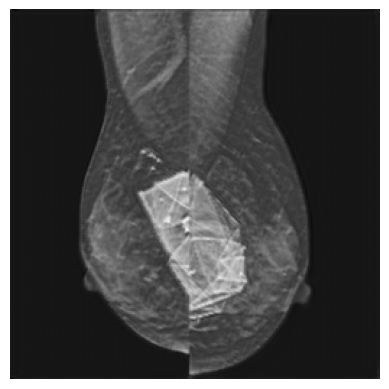} & 
    \includegraphics[width=0.1\linewidth]{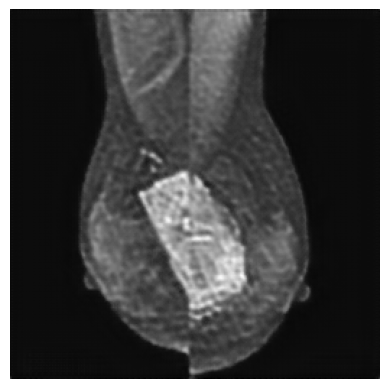} & 
    \includegraphics[width=0.1\linewidth]{images/recon_external_models/hybrid/external_implant_rap.png} \\
    \hline
    Biopsy Clip & 
    \includegraphics[width=0.1\linewidth]{images/recon_flag/external/mayo_bc.png} & 
    \includegraphics[width=0.1\linewidth]{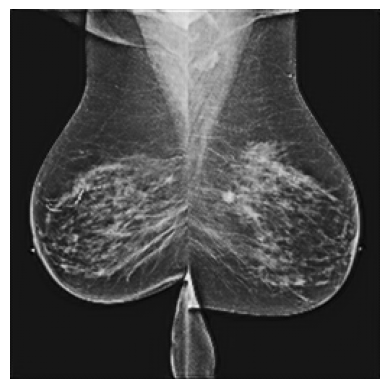} & 
    \includegraphics[width=0.1\linewidth]{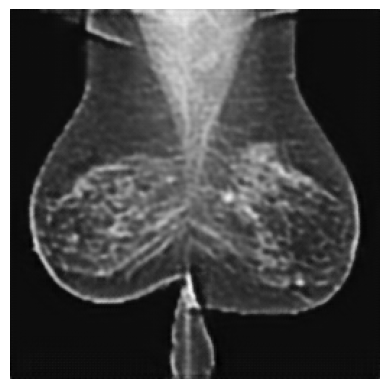} & 
    \includegraphics[width=0.1\linewidth]{images/recon_external_models/hybrid/external_bc.png} \\
    \hline
    Poor Quality & 
    \includegraphics[width=0.1\linewidth]{images/recon_flag/external/mayo_qi.png} & 
    \includegraphics[width=0.1\linewidth]{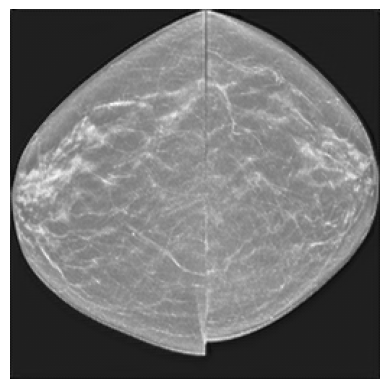} & 
    \includegraphics[width=0.1\linewidth]{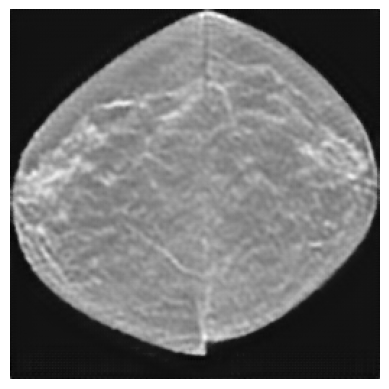} & 
    \includegraphics[width=0.1\linewidth]{images/recon_external_models/hybrid/external_qi.png} \\
    \hline
  \end{tabular}
  %\end{adjustbox}
  }
\end{table}
% \end{landscape}
\section{Discussion}\label{sec5}
Medical OOD detection poses great challenges due to the heterogeneity and unknown characteristics of medical data. 1) \emph{Mutations can happen.} Different from natural objects with fixed attributes, known diseases may progress to other mutated versions and generate anomalous data, e.g. breast cancer; 
2) \emph{Heterogeneous data is a big concern.} Medical images collected from different race groups can introduce heterogeneity, e.g. Asian patients have denser breast~\cite{del2007mammographic}; 3) \emph{Distribution shifting always exists.} Data scanned with different machines or at various sites may have intensity distribution shifting; 4) \emph{Data with defects is common.}  Medical images can be overexposed or scanned with incorrect positions/angles. Therefore, to train a novelty detector with only ID data available, learning high-quality ``normality" features is the fundamental step to identify the OOD samples during inference. Unfortunately, it still lacks an effective way to identify the difference for several datasets from the same medical domain. The main challenge lies in the inaccessibility to external medical datasets given the privacy concerns around sharing personally identifiable information~\cite{schutte2021overcoming}. Therefore, an efficient way of external dataset curation by identification of anomaly without sharing data is desired, particularly for AI model generalization. 

We proposed a novel hybrid architecture with transformer and CNN that can be trained in an unsupervised way to detect both intra- and inter-class anomalies from breast mammogram screening exams. The base hybrid architecture -TransBTS ~\cite{li2022transbtsv2}, resulted similar anomaly accuracy as the simple encoder based architecture. Compared to the existing SOTA, our primary contribution is to extend the hybrid architecture~\cite{li2022transbtsv2} with a discriminator branch and gradient reversal to support efficient detection of intra- and inter- class of anomaly samples. We also propose an synthetic ODD generation technique leveraging geometric augmentation to support unsupervised training of the model which can be extended for new use-cases. We trained the model using only ID data from RSNA Screening Mammography Breast Cancer Detection AI Challenge (2023) dataset and validate the model on OOD data from RSNA and unseen Mayo clinic external data. 

The proposed model outperformed encoder-based and GAN-based baselines which have been trained on the same ID data. Interestingly, it also outperformed the hybrid baselines and hybrid model with discriminator branch with significant margins, particularly for the external data which highlights the fact that addition of discriminator loss and penalizing the model with negative gradient help to learn fine-grain ``normality" features. Such training supports both intra- and inter-class anomaly detection for mammogram, even for small anomalies, such as presence of biopsy clip. With addition of each proposed component, we observed a constant improvement for ID reconstruction and worsening for OOD reconstruction quality (Fig ~\ref{fig:ssim}). 

The proposed hybrid pipeline can serve as a automated computational method for domain-specific quality checks and derive powerful and actionable insights. Once trained, the proposed anomaly detectors should be able to identify unknown anomalous patterns from an external large-scale mammogram screening dataset without ever seeing or having any prior knowledge of such any anomalous data examples in training. This quality makes the proposed pipeline particularly suitable for medical image dataset curation and AI model validation since exchanging healthcare data among institutions and manually identifying noisy or anomalous data are both extremely challenging in the current healthcare situation.

Experiments have several limitations. First, the model was only validated on a single external site with digital mammograms due the unavailability of the open-source mammogram images. EMBED~\cite{jeong2023emory} is another open-source digital mammogram dataset but a portion is already contained within the RSNA dataset which is used for the model training. Inclusion of that data as external could reflect anomaly performance with data leaking. Second, anomaly detectors were evaluated for detection of anomalous image characteristics and presence of foreign body, and not used for patient population drift detection which could be an important cause for performance drop. But we believe that considering population drift as anomalous sample is not a 'fair' definition, particularly for the minority sub-class.   

\section*{Funding}
Research reported in this paper was supported by NCI of the National Institutes of Health under award number U01 CA269264-01-1, Flexible NLP toolkit for automatic curation of outcomes for breast cancer patients .

\section*{Data statement}
RSNA Screening Mammography Breast Cancer Detection AI Challenge (2023) dataset is a publicly available dataset provided by RSNA. Mayo Clinic dataset is available upon proper data usage agreement.

\appendices

\section{HAND Network Details}\label{secA1}

Complete network details is provided in Table \ref{table:hand_network}.

\begin{table}[h!]
\centering
\caption{Network details of HAND with layers and output dimension.}
\label{table:hand_network}
\small  % Reduce font size
\resizebox{\columnwidth}{!}{%
\begin{tabular}{|c|c|c|c|}
\hline
\textbf{Stage} & \textbf{Block Name} & \textbf{Details} & \textbf{Output Size} \\ \hline
\multirow{2}{*}{Input} & \multicolumn{2}{c|}{--} & 1 $\times$ 256 $\times$ 256 \\ \cline{2-4}
                       & InitConv & Conv2d, Dropout & 4 $\times$ 256 $\times$ 256 \\ \hline
\multirow{7}{*}{\begin{tabular}[c]{@{}c@{}}CNN\\ Encoder\end{tabular}} & EnBlock1 & BN, ReLU, Conv2d & 4 $\times$ 256 $\times$ 256 \\ \cline{2-4}
                       & DownSample1 & Conv2d (kernel 3, stride 2) & 8 $\times$ 128 $\times$ 128 \\ \cline{2-4}
                       & EnBlock2 & BN, ReLU, Conv2d & 8 $\times$ 128 $\times$ 128 \\ \cline{2-4}
                       & DownSample2 & Conv2d (kernel 3, stride 2) & 16 $\times$ 64 $\times$ 64 \\ \cline{2-4}
                       & EnBlock3 & BN, ReLU, Conv2d & 16 $\times$ 64 $\times$ 64 \\ \cline{2-4}
                       & DownSample3 & Conv2d (kernel 3, stride 2) & 32 $\times$ 32 $\times$ 32 \\ \cline{2-4}
                       & EnBlock4 & BN, ReLU, Conv2d & 32 $\times$ 32 $\times$ 32 \\ \hline
\multirow{2}{*}{\begin{tabular}[c]{@{}c@{}}Transformer\\ Encoder\end{tabular}} & Linear Projection & Conv2d, Reshape & 128 $\times$ 256 (d $\times$ N) \\ \cline{2-4}
                       & Transformer & Transformer Layer $\times$ 4 & 128 $\times$ 256 (d $\times$ N) \\ \hline
\multirow{2}{*}{Discriminator} & MLP & Sigmoid & 1 \\ \cline{2-4}
                       & Feature Mapping & Reshape, Conv2d & 32 $\times$ 16 $\times$ 16 \\ \hline
\multirow{9}{*}{\begin{tabular}[c]{@{}c@{}}CNN\\ Decoder\end{tabular}} & DeBlock1 & Conv2d, BN, ReLU & 32 $\times$ 16 $\times$ 16 \\ \cline{2-4}
                       & UpSample1 & Conv2d, DeConv2d & 16 $\times$ 32 $\times$ 32 \\ \cline{2-4}
                       & UpSample2 & Conv2d, DeConv2d & 16 $\times$ 64 $\times$ 64 \\ \cline{2-4}
                       & UpSample3 & BN, ReLU, Conv2d & 8 $\times$ 128 $\times$ 128 \\ \cline{2-4}
                       & DeBlock2 & Conv2d, DeConv, Conv2d & 8 $\times$ 128 $\times$ 128 \\ \cline{2-4}
                       & DeBlock2d & BN, ReLU, Conv2d & 4 $\times$ 256 $\times$ 256 \\ \cline{2-4}
                       & UpSample4 & Conv2d, DeConv2d & 4 $\times$ 256 $\times$ 256 \\ \cline{2-4}
                       & EndConv & Conv2d, Softmax & 1 $\times$ 256 $\times$ 256 \\ \hline
\end{tabular}
}
\normalsize  % Return to normal size after the table
\end{table}

\section{Data Pre-processing}\label{secA2}
\begin{figure}[htb!]
\centering
\begin{subfigure}[]{}
\includegraphics[width=0.47\linewidth] {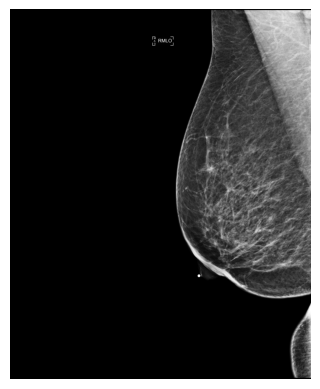}
\end{subfigure}
\begin{subfigure}[]{}
\includegraphics[width=0.47\linewidth]
{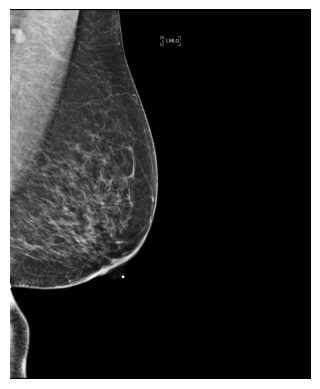}
\end{subfigure}
\begin{subfigure}[]{}
\includegraphics[width=0.7\linewidth]{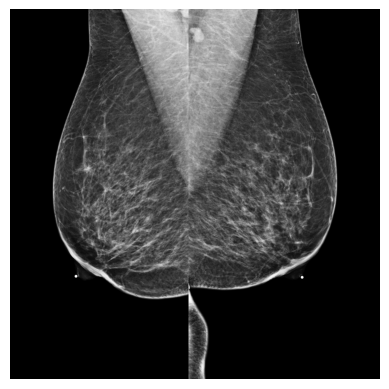}
\end{subfigure}
\caption{Examples of pre-process mammograms (crop and stitch right and left MLO view under same study ID) - (a) Original MLO right view. (b) Original MLO left view. (c) Cropped and stitched MLO right and MLO left under same study ID.}
\label{fig:process_mammo}
\end{figure}

Mammogram images used in this study have been processed in the following steps. First, from original DICOM files, an open source ImageExtractor - Niffler\footnote{https://github.com/Emory-HITI/Niffler}, is applied to do proper windowing and also extract comprehensive DICOM metadata as a single CSV file. The metadata is merged with clinical labels (presence of implant, biopsy, cancer etc). We have drop duplicate images of same laterality (L/R) and view (CC/MLO) under same study ID according to latest acquisition time which in results  47,688 records, with MLO and CC view each contains 23,826 records. 684 implant records, distributed equally between MLO and CC. Inversion of a image been performed if `MONOCHROME' dicom metadata tag is 1 (white background with black tissue). After the intensity based pre-processing, we cropped the breast tissue from background using pre-defined thresholds and we have decided to stitch right and left MLO mammogram images together to represent the study. Fig. \ref{fig:process_mammo}.c. shows an example of stitched right (Fig. \ref{fig:process_mammo}.a.) and left (Fig. \ref{fig:process_mammo}.b.) MLO views under the same study ID. If there is one side (right or left) missing for the same study ID and same view, we have dropped these cases. Resulting in 22,778 records of stitched mammograms, with MLO and CC views distributed equally. 

Even though labeled implant cases have been dropped for training dataset, there are still a good amount of samples in the training datasets contains implants and other noises. Thus, to facilitate strict `normality learning', we decided to reduce noisy samples by dropping pixel intensity far from normal pattern for both MLO (4,934) and CC (3,845) views, which results in total 13,999 samples. Then, a manual review is done to reduce more noisy samples. Finally, 11,722 samples are used in training dataset. 

\section*{References}

\def\refname{\vadjust{\vspace*{-2.5em}}} 

\bibliographystyle{IEEEtran}
\bibliography{IEEE-TJ-color-latex-template/main}

\end{document}